\documentclass[sigplan,10pt]{acmart}
\usepackage{diagbox}
\usepackage[normalem]{ulem}
\usepackage[export]{adjustbox}
\usepackage{listings}
\usepackage{soul}
\usepackage{mdframed}
\mdfdefinestyle{revstyle}{linecolor=yellow, linewidth=2pt, innertopmargin=2pt, innerbottommargin=2pt, innerleftmargin=2pt, innerrightmargin=2pt}

\soulregister\cite7
\soulregister\ref7
\soulregister\pageref7

\newcommand{\eg}{e.g.,\xspace}
\newcommand{\ie}{i.e.,\xspace}

\newcommand{\xname}{NearPM\xspace}
\newcommand{\pname}{PPO\xspace}

\usepackage[normalem]{ulem}
\usepackage{xspace}
\usepackage{xcolor}
\usepackage{fancyhdr}
\usepackage{microtype}
\usepackage{etoolbox}
 \usepackage{booktabs} 
\usepackage{multirow}
\usepackage{footnote}
\makesavenoteenv{tabular}
\makesavenoteenv{table}
\usepackage[percent]{overpic}
\usepackage{subcaption}
\usepackage{enumitem}
 \usepackage{algorithm2e}
\usepackage{float}
\usepackage[normalem]{ulem}
\usepackage{wrapfig}
\usepackage{comment}
\usepackage[noabbrev,capitalise]{cleveref}

\usepackage{color,soul}
\usepackage[export]{adjustbox}
\usepackage{amsmath,amsfonts}

\usepackage{array}
\newcolumntype{L}[1]{>{\raggedright\let\newline\\\arraybackslash\hspace{0pt}}m{#1}}
\newcolumntype{C}[1]{>{\centering\let\newline\\\arraybackslash\hspace{0pt}}m{#1}}
\newcolumntype{R}[1]{>{\raggedleft\let\newline\\\arraybackslash\hspace{0pt}}m{#1}}

 \PassOptionsToPackage{hyphens}{url}

\newcommand{\squishlist}{
   \begin{list}{$\bullet$}
    { \setlength{\itemsep}{0pt}      \setlength{\parsep}{3pt}
      \setlength{\topsep}{3pt}       \setlength{\partopsep}{0pt}
      \setlength{\leftmargin}{1.5em} \setlength{\labelwidth}{1em}
      \setlength{\labelsep}{0.5em} } }

\newcommand{\squishend}{
    \end{list}  }
    
\usepackage{pifont}
\newcommand{\circled}[1]{\ding{\numexpr#1+201}}
\newcommand{\circledtext}[1]{\raisebox{.5pt}{\textcircled{\raisebox{-0.9pt} {\footnotesize #1}}}}

\usepackage{graphicx}
\newcommand\sbullet[1][.5]{\mathbin{\vcenter{\hbox{\scalebox{#1}{$\bullet$}}}}}

\makeatletter
\def\SOUL@hlpreamble{%
\setul{}{2.2ex}
\let\SOUL@stcolor\SOUL@hlcolor
\SOUL@stpreamble
}
\makeatother
\soulregister\xname7
\soulregister\pname7
\soulregister\cite7
\soulregister\ref7
\soulregister\cref7
\soulregister\pageref7
\soulregister\eg7
\soulregister\ie7
\soulregister\texttt7
\soulregister\circled7
\soulregister\encircle7
\soulregister\uline7
\soulregister\circled7
\soulregister\circledtext7
\acmYear{2023}\copyrightyear{2023}
\setcopyright{rightsretained}
\acmConference[EuroSys '23]{Eighteenth European Conference on Computer Systems}{May 8--12, 2023}{Rome, Italy}
\acmBooktitle{Eighteenth European Conference on Computer Systems (EuroSys '23), May 8--12, 2023, Rome, Italy}
\acmPrice{}
\acmDOI{10.1145/3552326.3587456}
\acmISBN{978-1-4503-9487-1/23/05}



\begin{document}

\title{NearPM: A Near-Data Processing System for Storage-Class Applications}


\author{Yasas Seneviratne}
\affiliation{%
  \institution{University of Virginia}
  \country{United States}}
\email{yasas@virginia.edu}

\author{Korakit Seemakhupt}
\affiliation{%
  \institution{University of Virginia}
  \country{United States}}
\email{korakit@virginia.edu}

\author{Sihang Liu}
\authornote{This author contributed to this work at the University of Virginia.}
\affiliation{%
  \institution{University of Waterloo}
  \country{Canada}}
\email{sihangliu@uwaterloo.ca}

\author{Samira Khan}
\affiliation{%
  \institution{University of Virginia}
  \country{United States}}
\email{samirakhan@virginia.edu}

\date{}

\begin{abstract}
Persistent Memory (PM) technologies enable both fast memory access and recovery in case of a failure. 
To ensure crash-consistent behavior, programs need to enforce persist ordering and employ mechanisms that introduce additional data movements such as logging, checkpointing, and shadow-paging. 
The emerging near-data processing (NDP) architectures can effectively reduce this overhead.
In this work, we propose \xname{}, a near-data processor that accelerates common, primitive operations that are crucial to crash consistency.
Using these primitives, \xname{} accelerates commonly-used crash-consistency mechanisms.
\xname{} further reduces the synchronization overheads between the NDP and the CPU by handling ordering near memory.
We propose Partitioned Persist Ordering (PPO) that ensures a correct persist ordering between CPU and NDP devices, as well as among multiple NDP devices. 
We prototype \xname{} on an FPGA platform. 
\xname{} executes the data-intensive operations of crash-consistency mechanisms with correct ordering guarantees, while the rest of the program runs on the CPU.
We evaluate nine PM workloads, each implemented in three crash consistency mechanisms:  logging, checkpointing, and shadow paging. 
Overall, \xname{} achieves $4.3-9.8\times$ speedup in the NDP-offloaded operations and $1.22-1.35\times$ speedup in the whole applications.
\end{abstract}
\maketitle

\thispagestyle{empty} 
\pagestyle{empty}

\section{Introduction}

Persistent main memory (PM) technologies offer both high performance and data persistence. 
For example, Optane PM \cite{optane} can be accessed through the DDR interface. 
Other alternative PM systems \cite{intel_pmem_cxl, lenovo_pemm_cxl}  are also being developed for the upcoming PCIe-based Compute Express Link (CXL) standard \cite{cxl}.  
Unlike conventional storage devices (e.g., HDD and SSD), these PM systems enable applications to perform direct access to PM, without going through the file system intermediaries.
Thus, PM-optimized applications can benefit from a faster data path. 
These opportunities have inspired research on developing and deploying PM \cite{shanbhag2020IDM, gill2019arx, lee19_recipe, izraelevitz19_dcpmm, google_nvm, amazon_nvm, chen15_vldb, arulraj2018_vldb}. 
However, a new challenge arises---without the file system, it is now up to the  applications to manage the recovery of persistent data. 
In case of a failure (e.g., a system crash or power outage), applications that directly access PM need to ensure that the persistent data is maintained in a recoverable state. We call this property the crash consistency guarantee.

There have been myriad solutions that provide crash consistency guarantees for PM-based applications. 
For example, the undo-logging approach makes a backup of the existing data to PM before updates \cite{chakrabarti14_oopsla, chatzistergiou15_pvldb, coburn13_sosp,  coburn11_asplos,dulloor14_eurosys,pmdk, kolli16_asplos, pmem-memcached, gogte18_pldi}; the checkpointing method periodically makes a snapshot of persistent data to keep a consistent, recoverable copy \cite{giles17_ismm, kannan13_ipdps, bailey2013exploring, ongaro11_sosp, ren15_micro,fernando2016_hipc}; the shadow-paging mechanism redirects writes to a shadow memory and changes page reference at commit \cite{hsu17_eurosys, wu2020_pldi,ni18_hotstorage,ni2019_micro}.
However, these mechanisms come with a performance cost. 
First, crash consistency guarantees require writes to become persistent in a specific order, introducing additional ordering constraints.
For example, the undo-logging mechanism backs up the to-be-updated persistent data \emph{before} performing the update.
Therefore, crash consistency mechanisms introduce additional stalls to the program execution, as they need to maintain a correct persist ordering \cite{kolli16_asplos,volos11_asplos,xu16_fast,pelley14_isca, nalli17_asplos}.  
Second, these crash consistency mechanisms usually make extra copies of data \cite{condit09_sosp, chatzistergiou15_pvldb,coburn13_sosp,hsu17_eurosys,pmdk,izraelevitz16_asplos} in order to recover in case of a failure. 
Such data movement introduces additional memory bandwidth utilization. 
Combining these two performance bottlenecks, crash-consistency mechanisms can place extra data-intensive operations on the critical path.

Near-data processing (NDP) is an emerging computer architecture design that has the potential to mitigate these overheads.
By bringing computation closer to data, NDP can reduce the data movement between memory and processor (\eg CPU)~\cite{fernandez2020_ICCD,hsieh2016_sigarch,kim2016_sigarch,lockerman20_livia}. 
In particular, as the new CXL standard \cite{cxl} is around the corner, more opportunities for processing closer to PM devices are opening.

In this work, we propose \xname{}, a near-data processing system for PM-based storage-class applications. 
\xname{} is integrated into PM memory devices, with access to the full memory of that device.
However, it is not straightforward to accelerate PM programs as each program may follow a different crash consistency mechanism---simply adding hardware acceleration logic for each mechanism is impractical. 
We observe that different crash consistency mechanisms share common primitive operations. 
For example, undo-logging copies the original persistent data to a log and updates the persistent data in-place. 
On the other hand, redo-logging first updates logs and then copies the logs back to the original location. 
These two crash consistency techniques both copy data to a log and update PM. 
Thus, by supporting common accelerable primitives in the hardware and using them as building blocks, \xname{} is capable of accelerating various crash consistency mechanisms.

Accelerating primitive operations speeds up the execution but the offloaded execution still needs to satisfy the ordering constraints. 
Naively, one could enforce the same ordering guarantees as the original CPU-based program but at the cost of excessive CPU-NDP synchronization. 
To overcome the synchronization overheads, our approach is to handle ordering near memory, enabling NDP execution to overlap with CPU procedures.

However, handling ordering near memory has new challenges.
Ordering guarantees must be satisfied even though the execution is partitioned between the CPU and the \xname{}.
For example, when offloading undo-logging to \xname{}, the log must be persisted by \xname{} to memory before the CPU performs an in-place update.
Furthermore, the program execution is not only partitioned between CPU and \xname{} but may also happen across multiple \xname{} devices.
For example, two interleaved \xname{} devices may both hold a fraction of the persistent data and the execution flows on both devices can be out of sync. 
Consider again the undo-log example, assume that some object is interleaved among two devices and that one device has the update committed but the other is still backing up data to the log. 
When a failure occurs, recovery might keep the updates in one device but roll back updates on another, leading to inconsistencies.

To overcome the challenges we define Partitioned Persist Ordering (\pname{}) for correct offloaded execution in NDP-enabled PM systems. 
PPO defines persist ordering in two scenarios. 
The first scenario is the order between the CPU and \xname{} operations. 
Naively, one would enforce strict persist ordering between CPU and \xname{} to provide the same crash consistency guarantees as the original program. 
However, this approach offsets the performance benefits of NDP. 
We observe that persists from \xname{} to memory addresses that are managed by \xname{} only but not shared with the CPU, such as logs, do not need to follow the original ordering constraints. 
Therefore, such a relaxed persist ordering mitigates CPU-side stalling and further allows \xname{} operations to execute in parallel. 
For example, without back-and-forth synchronization, the CPU-side procedure can issue multiple independent logging operations to \xname{} and have them executed in parallel. 

The second scenario is the synchronization among multiple \xname{} devices. 
Naively, frequent synchronizations among \xname{} devices after every offloaded crash consistency operation keep them at the same pace and ensure their completion before committing updates. 
Similar to the case of CPU-NDP ordering, such a naive solution degrades performance benefits from NDP. 
Because NDP-managed memory, such as logs, would not be accessed nor exposed to the CPU-side  procedure unless recovery happens. 
Therefore, it is possible to delay the synchronization and move it off the critical path of the program execution.
As long as data needed for recovery remains intact until the completion of a series of PM operations (e.g., commit in a PM transaction), data can still recover successfully.

\begin{figure*}
    \centering
    \begin{subfigure}[t]{0.22\linewidth}
        \centering
        \includegraphics[height=1in]{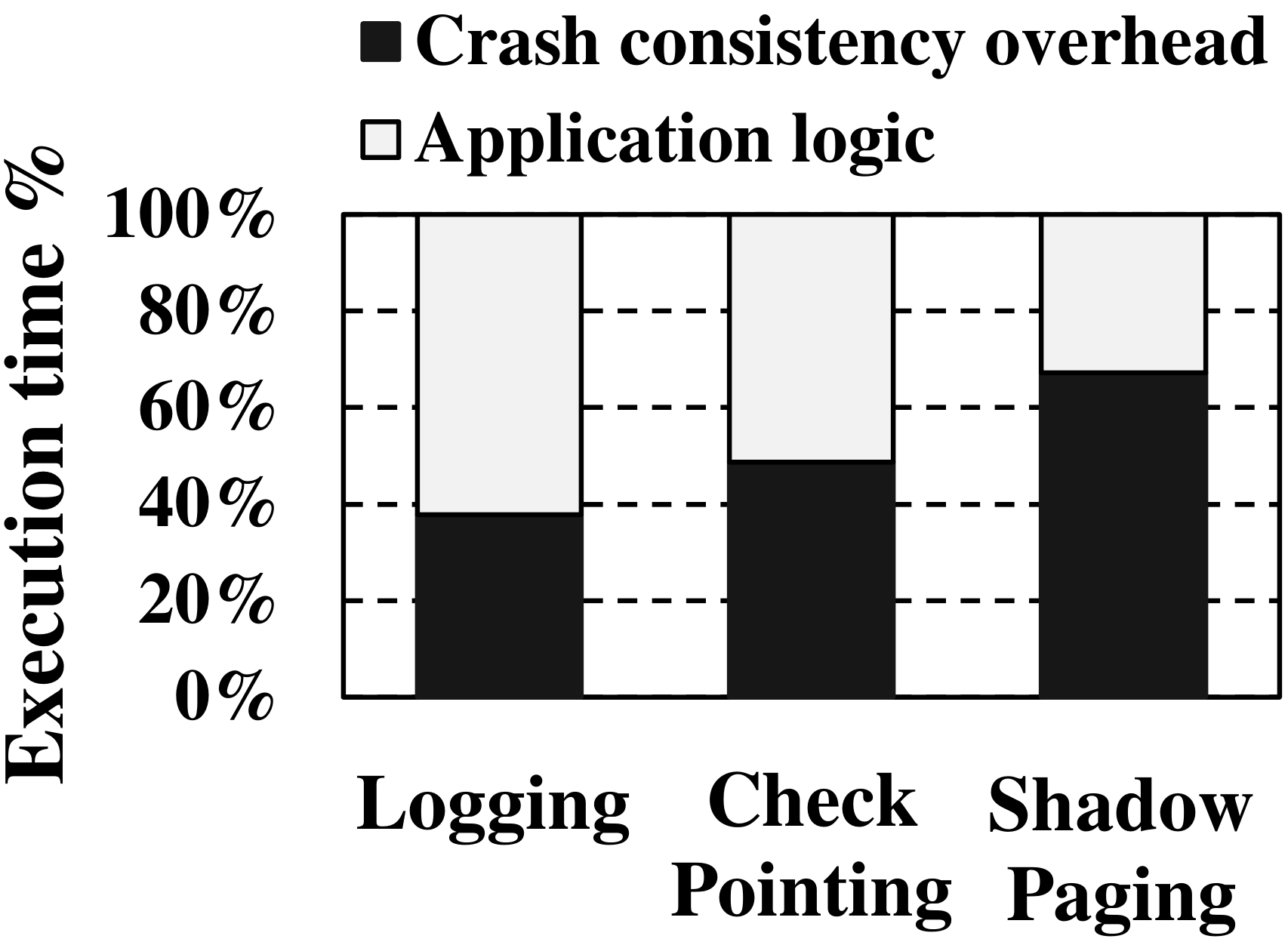}
        \caption{Crash consistency overhead}
        \label{fig:crashconsistency_overhead}
    \end{subfigure}
    \hspace{-0.1in}
    \begin{subfigure}[t]{0.28\linewidth}
        \includegraphics[height=1.01in]{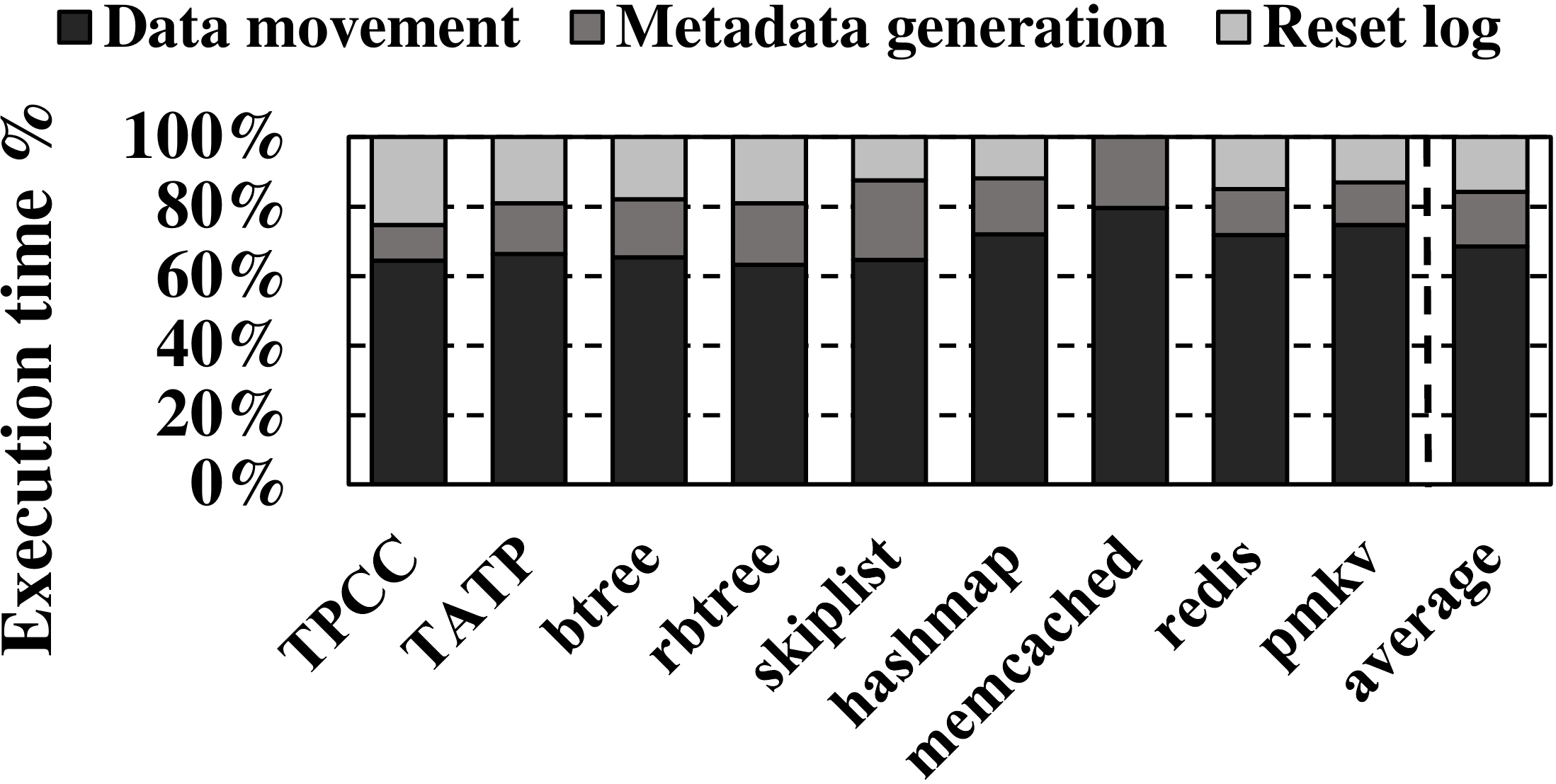}
        \caption{Logging}
        \label{fig:logbreakdown}
    \end{subfigure}
    \begin{subfigure}[t]{0.25\linewidth}
        \centering
        \includegraphics[height=1.02in]{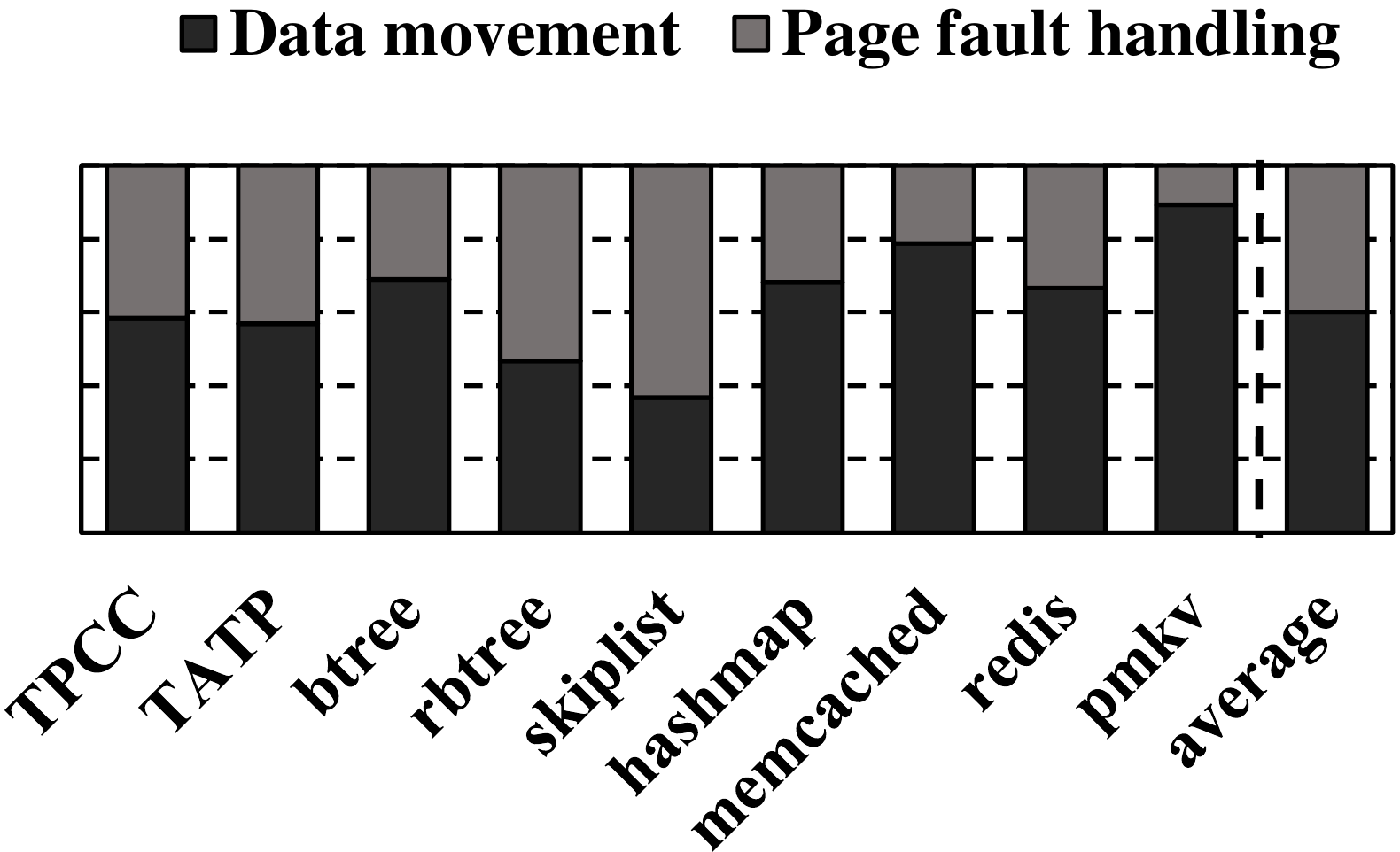}
        \caption{Checkpointing}\
        \label{fig:checkpointbreakdown}
    \end{subfigure}
    \hspace{-0.1in}
    \begin{subfigure}[t]{0.25\linewidth}
        \centering
        \includegraphics[height=1.02in]{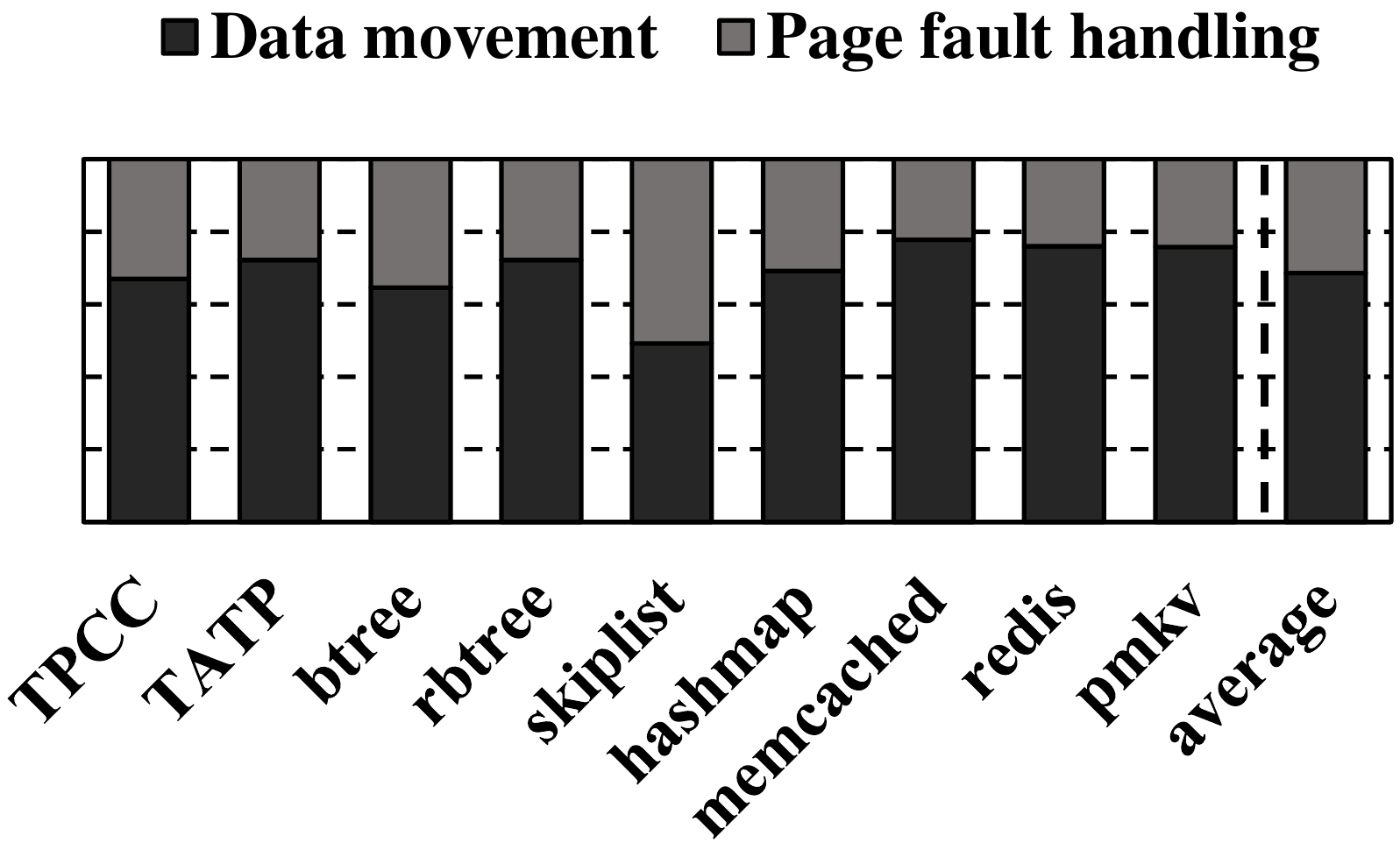}
        \caption{Shadow paging}\
        \label{fig:shadowbreakdown}
    \end{subfigure}
    \vspace{-0.15in}
    \caption{Crash consistency overheads (a) and the breakdown in logging, checkpointing, and shadow paging (b--d).}
\end{figure*}

With the key insights above, we provide primitive operations in crash consistency mechanisms to cover a wide range of PM-base storage-class applications and introduce \pname{} to mitigate overheads stemming from the crash consistency guarantees in partitioned execution of NDP systems.
To evaluate our design, we prototype \xname{} on an FPGA platform that connects to the host CPU through PCIe. 
\footnote{The software and hardware implementation of \xname{} are available at \url{https://github.com/Systems-ShiftLab/NearPMSW} and \url{https://github.com/Systems-ShiftLab/NearPMHW}, respectively.}
We emulate PM using the on-board memory on the FPGA platform, allowing the CPU and the NDP device on the FPGA to access memory using loads/stores. 
We evaluate nine PM-optimized workloads with two \xname{} devices. Each workload has implementations for logging, checkpointing, and shadow paging.
The main contributions are the following:
\begin{itemize}[leftmargin=*]
    \item We propose a near-data processing architecture that accelerates crash consistency mechanisms, by supporting common accelerable primitives.
    \item We define \emph{partitioned persist ordering} (\pname{}) that defines the persist ordering in NDP systems. 
    \pname{} ensures correctness and performance when execution is partitioned among the CPU and multiple NDP devices. 
    \item We prototype \xname{} using an FPGA platform. Our evaluation shows that \xname{} reduces the crash consistency overhead by $6.97\times$, $4.26\times$, and $9.76\times$ in logging, checkpointing, and shadow-paging-based programs, compared to the CPU-only baseline.
    In terms of the end-to-end performance in the whole program, it achieves $1.35\times$, $1.22\times$, and $1.33\times$ speedup over the baseline, respectively.
\end{itemize}

\section{Background and Motivation} \label{sec:background}

\begin{figure}
  \centering
 \includegraphics[width=\linewidth]{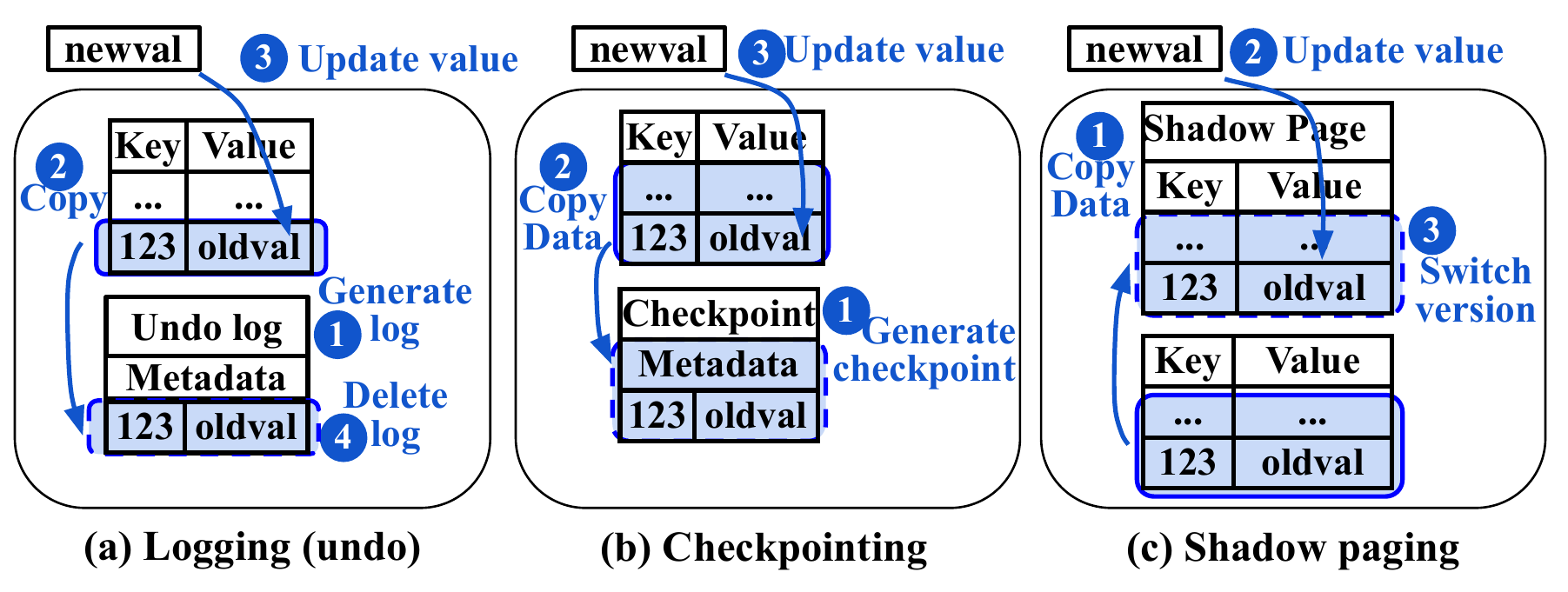}
  \caption{Procedures in crash consistency mechanisms.}
  \label{fig:crashcons}
\end{figure}

\subsection{Crash Consistency and PM Programming}\label{sec:background_crashconsistency}

Persistent memory technologies (PM) feature high performance, data persistence, and byte-addressable direct access to persistent data bypassing the file system.
Intel Optane PM~\cite{optane} is a memory module that shares the memory bus with DRAM modules; other upcoming PM technologies \cite{bhardwaj2022cache, intel_pmem_cxl, lenovo_pemm_cxl} will be on the PCIe bus but enable direct access via the Compute Express Link (CXL) \cite{cxl}.

Direct access to persistent data reduces overhead on the data path, but at the same time moves the burden of managing data recovery to applications.
We refer to the ability to restore persistent data after a failure (\eg a power outage or system crash) as the \emph{crash consistency guarantee}.
Past research on crash-consistent programming has proposed various mechanisms for the crash consistency guarantee, such as undo-logging redo-logging \cite{chakrabarti14_oopsla,chatzistergiou15_pvldb,hsu17_eurosys,pmdk,kolli16_asplos}, checkpointing \cite{fernando2016_hipc,kannan13_ipdps}, and shadow paging \cite{condit09_sosp,ni18_hotstorage}.

Logging replicates persistent data to a separate location (e.g., an undo or a redo log) before updating the persistent state. 
As shown in \cref{fig:crashcons}a, undo-logging makes a fine-grained snapshot of the original data in a log, before persisting the in-place updates; it deletes the log only after the latter completes.
Similarly, redo-logging redirects each update to a log, and applies the updates in-place only after the log has become persistent.
Checkpointing (\cref{fig:crashcons}b) maintains a coarse-grained snapshot of persistent locations prior to updates.  
Shadow paging first redirects the update to a newly allocated page and then changes the references to the original page to the new version (\cref{fig:crashcons}c).

These mechanisms introduce performance overheads.
The main overhead is due to intensive data movement. 
\cref{fig:crashconsistency_overhead} shows that logging, checkpointing, and shadow paging mechanisms take up 37.7\%, 48.6\%, and 67.2\% of the execution time, respectively (methodology in \cref{subsec:methodology}).
Figures \ref{fig:logbreakdown}, \ref{fig:checkpointbreakdown}, and \ref{fig:shadowbreakdown} further break down the crash consistency overhead, showing that 68.9\%, 60.4\%, and 70.5\% of the overhead is from data movement in these crash consistency mechanisms, respectively.
Thus, there is a huge opportunity for acceleration. 

\subsection{Near-Data Processing (NDP)}

\begin{figure}
  \centering
 \includegraphics[width=\linewidth]{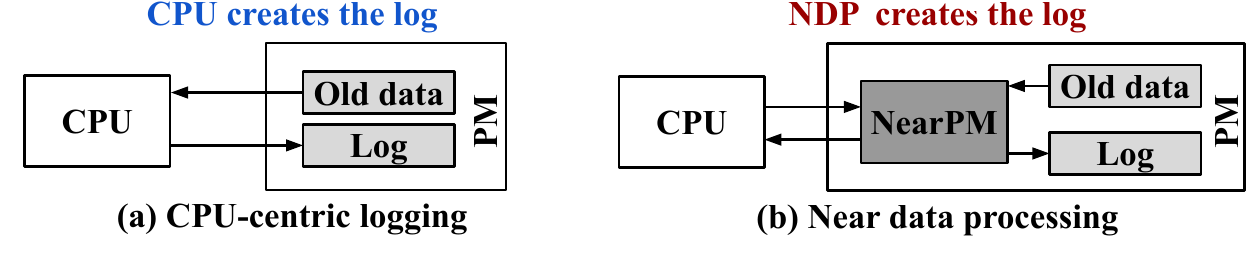}
  \caption{CPU- and NDP-based log generation.}\
  \label{fig:neardata}
\end{figure}

In traditional systems, the CPU is in charge of manipulating data. 
For instance, to create a copy of data in memory (as shown in \cref{fig:neardata}a), the CPU needs to fetch data through the cache hierarchy and write it to another CPU-manged memory location, leading to a high data movement overhead. 
To mitigate data movement overheads, researchers have introduced the paradigm of near-data processing (NDP) that places computation closer to the data \cite{ahn2015_ISCA2,fernandez2020_ICCD,gao2015_PACT,gao2016_HPCA,kim2018_BMC,mutlu2020modern,singh2019_DAC,zhan16_micro}.

NDP is well-suited for applications or code regions that are memory-intensive and feature high parallelism. 
Therefore, it has the potential to mitigate the overhead of memory-intensive operations involved in crash consistency mechanisms.
The existing and upcoming PM devices are also capable of hosting computation near or inside the PM device. 
For example, PCIe-based PM devices can integrate compression, query processing, and data movement logic \cite{bhardwaj2022cache,ahn2022enabling}; even the more compact PM devices, such as Optane DIMMs, already integrates controllers for data-intensive tasks \cite{wang20_micro} such as encryption, which can be extended to NDP.
\cref{fig:neardata}b illustrates how an NDP unit copies data to a log without going through the CPU.

\subsection{Challenges of NDP Crash Consistency}\label{sec:challenges}

\begin{figure}
  \centering
 \includegraphics[width=1.0\linewidth]{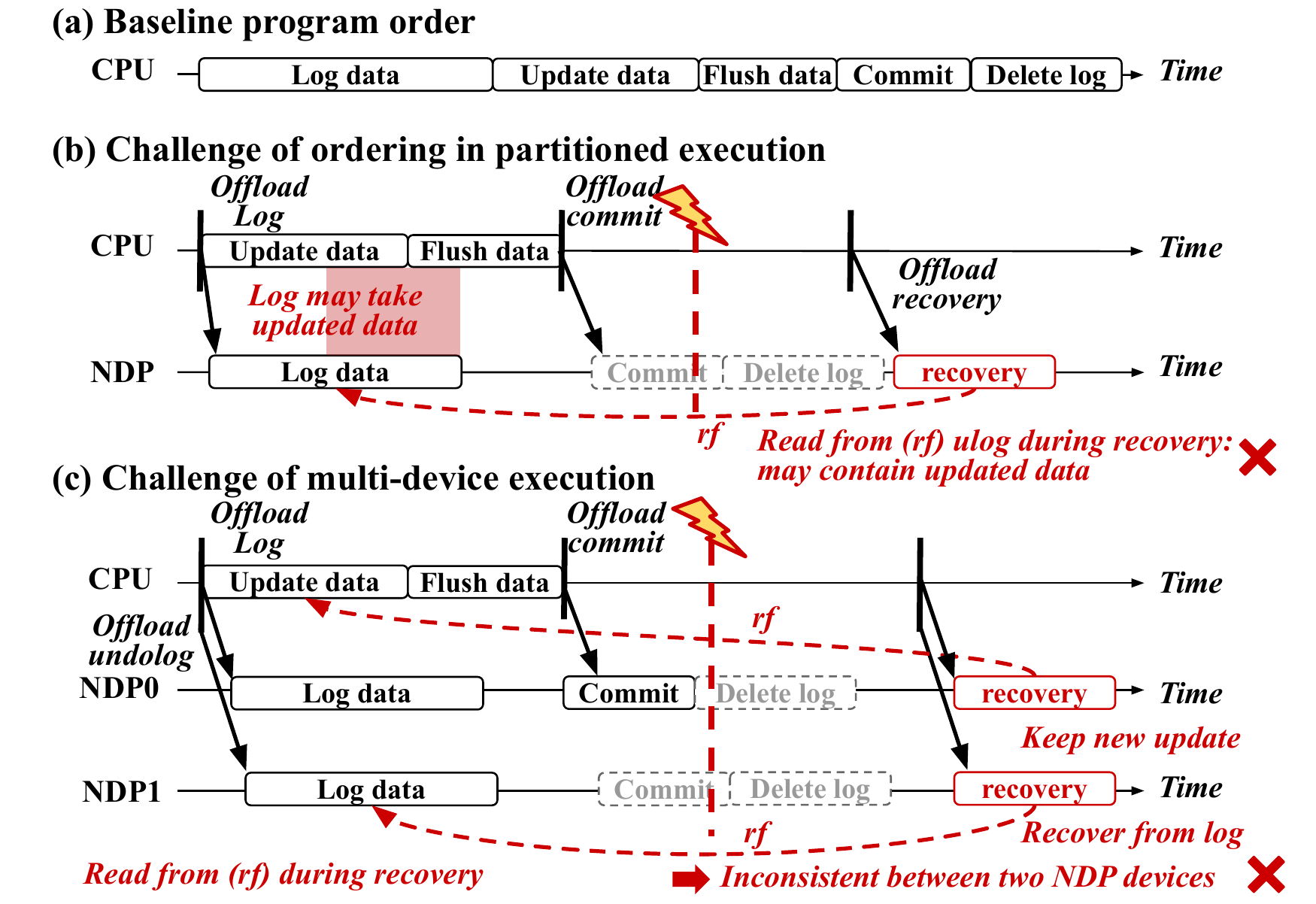}
  \caption{Challenges of ordering in partitioned execution}\
  \label{fig:orderchallenge}
\end{figure}

In this section, we discuss the challenges in supporting NDP processing for crash consistency operations. 

\paragraph*{Support for different crash consistency mechanisms.}
Although NDP can effectively accelerate memory-intensive procedures, NDP units are highly specialized.
As a result, it is challenging to accelerate general programs, as programmers first need to identify  NDP-friendly code regions and convert them based on NDP-accelerated hardware primitives.

As explained in \cref{sec:background_crashconsistency}, there is a diversity of crash consistency mechanisms, such as undo- and redo-logging, checkpointing, and shadow paging. 
Supporting every single one with its own dedicated acceleration logic is not realistic.  
Therefore, it is necessary to find a common ground in order to integrate NDP into a practical system.
We will discuss our high-level ideas of NDP acceleration for crash consistency operations in \cref{subsec:acc}.

\paragraph*{Ensuring persist ordering near memory.}
As shown in \cref{fig:crashcons}, crash consistency mechanisms enforce persist ordering. 
When offloading computation to an NDP-enabled PM device, program execution becomes \emph{partitioned} between the CPU and the PM device.
\cref{fig:orderchallenge} shows how a conventional CPU-centric system strictly orders undo-logging. 
\cref{fig:orderchallenge}b offloads undo-logging to an NDP, while the other steps remain executed on the CPU.
Such a partitioned execution breaks the ordering guarantees: the CPU concurrently  persists to PM, while the NDP unit is creating an undo log.
In case of a failure (as indicated by the red line), as the update was not committed, recovery attempts to read from (as indicated by the rf edge) the undo log.
Due to incorrect ordering, the undo log might contain already updated data, leading to an inconsistent recovery.

In addition, the execution may become partitioned among multiple NDP devices. 
\cref{fig:orderchallenge}c shows a scenario where two PM devices interleave. As such, a PM object can span both devices. 
The NDP units on the two devices operate on this partitioned object. 
However, without synchronizing the execution between them, the offloaded execution might progress at a different pace on both devices, a failure indicated by the red line, one PM device NDP0 has committed the update, but the other (NDP1) has not. 
As a result, during failure recovery, NDP0 maintains the in-place updates, as they were committed prior to failure, while NDP1 reads from the old copy in the log.
Thus, the recovered data is inconsistent: partly from the original version and partly from the updated version. 
We will discuss our high-level ideas that ensure correct persist ordering between CPU and NDP devices, and among NDP devices in \cref{subsec:persist_ordering}.

\section{High-level Ideas}

In this section, we will first discuss our high-level ideas of accelerating crash consistency operations with NDP, and then a persist ordering for partitioned execution among CPU and NDP devices. 

\subsection{Acceleration for Crash Consistency Operations}
\label{subsec:acc}

\begin{table}[t]
\aboverulesep = 0.22em
\belowrulesep = 0.22em

\caption{Evaluated crash consistency mechanisms.}
\label{tab:commonops}
\setlength{\tabcolsep}{3pt}
    \centering
    \footnotesize
    \begin{tabular}{L{3.8cm}L{4.3cm}}
        \toprule
        \textbf{Crash consistency mechanism} & \textbf{Common operations} \\
        \midrule
        Logging (undo) \cite{chakrabarti14_oopsla, chatzistergiou15_pvldb, coburn13_sosp,  coburn11_asplos,dulloor14_eurosys,pmdk, kolli16_asplos, pmem-memcached, gogte18_pldi}  & \multirow{4}{*}{\shortstack[l]{allocate, generate metadata,\\copy data, delete log, commit}} \\
        \cmidrule{1-1}
        Logging (redo) \cite{volos11_asplos,mohan92, giles15_msst,wu94_asplos} &  \\ 
        \cmidrule{1-1}
        Logging (undo+redo) \cite{pmdk,coburn13_sosp} &  \\ 
        \midrule
        Checkpointing \cite{fernando2016_hipc,giles17_ismm, kannan13_ipdps, bailey2013exploring, ongaro11_sosp, volos11_asplos,ren15_micro} & allocate, generate metadata, copy data\\
        \midrule
        Shadow paging \cite{hsu17_eurosys, wu2020_pldi,ni18_hotstorage,ni2019_micro} &  allocate, copy data, switch page \\ 
        \bottomrule
    \end{tabular}
   \vspace{-0.15in}
\end{table}


Our first insight is to divide the different crash consistency operations into smaller primitives, as shown in Table \ref{tab:commonops}. 
We observe that different crash consistency techniques consist of common accelerable routines.
For example, undo- and redo-logging both contain the following primitive operations: generate metadata (\eg object ID, commit status, offset in PMDK  \cite{pmdk} logs), copy data, and delete metadata. 
Therefore, our first key idea is to identify accelerable primitives in existing crash consistency mechanisms and implement NDP logic for these primitives in hardware near memory.
We use the term \emph{NDP procedure} to describe the execution of a series of consecutive NDP primitives that correspond to a crash consistency operation.

\cref{fig:undo_decoupled}a and \cref{fig:undo_decoupled}b compare executions of two undo-logging procedures with and without \xname{}. 
\cref{fig:undo_decoupled}a, everything goes through the CPU.
All operations are ordered sequentially (even when independent operations Update A and Log B), and the memory-intensive operations creating and deleting the logs use the CPU.
In \cref{fig:undo_decoupled}b accelerable primitives executed near-data.
For example, copying data to Log A and Log B will take a shorter amount of time because they are NDP-friendly.
Because persistent ordering is handled by the CPU when executing operations near memory, NDP devices need to synchronize with the CPU.
Furthermore, the CPU might remain idle while the NDP operations are completed.
This issue leads to our second key idea which is handling ordering near memory.

Handling ordering near memory removes the overhead of synchronization between \xname{} and the CPU.
In addition, this approach allows for more relaxed persistency, allowing the CPU and \xname{} to execute in parallel.
\cref{fig:undo_decoupled}c shows the benefit of handling ordering crash-consistent programs from the NDP side. 
Ordering PM programs near memory is not free because of the partitioned nature of the execution.

\subsection{Persistent Ordering for Partitioned Execution}
\label{subsec:persist_ordering}

Executing crash consistency operations near memory partitions the program execution. 
To ensure correct execution and failure-recovery, we propose \emph{partitioned persist ordering} (\pname{}) that ensures persist ordering between CPU and NDP, as well as among different NDP devices.

\begin{figure}
  \centering
 \includegraphics[width=1.0\linewidth]{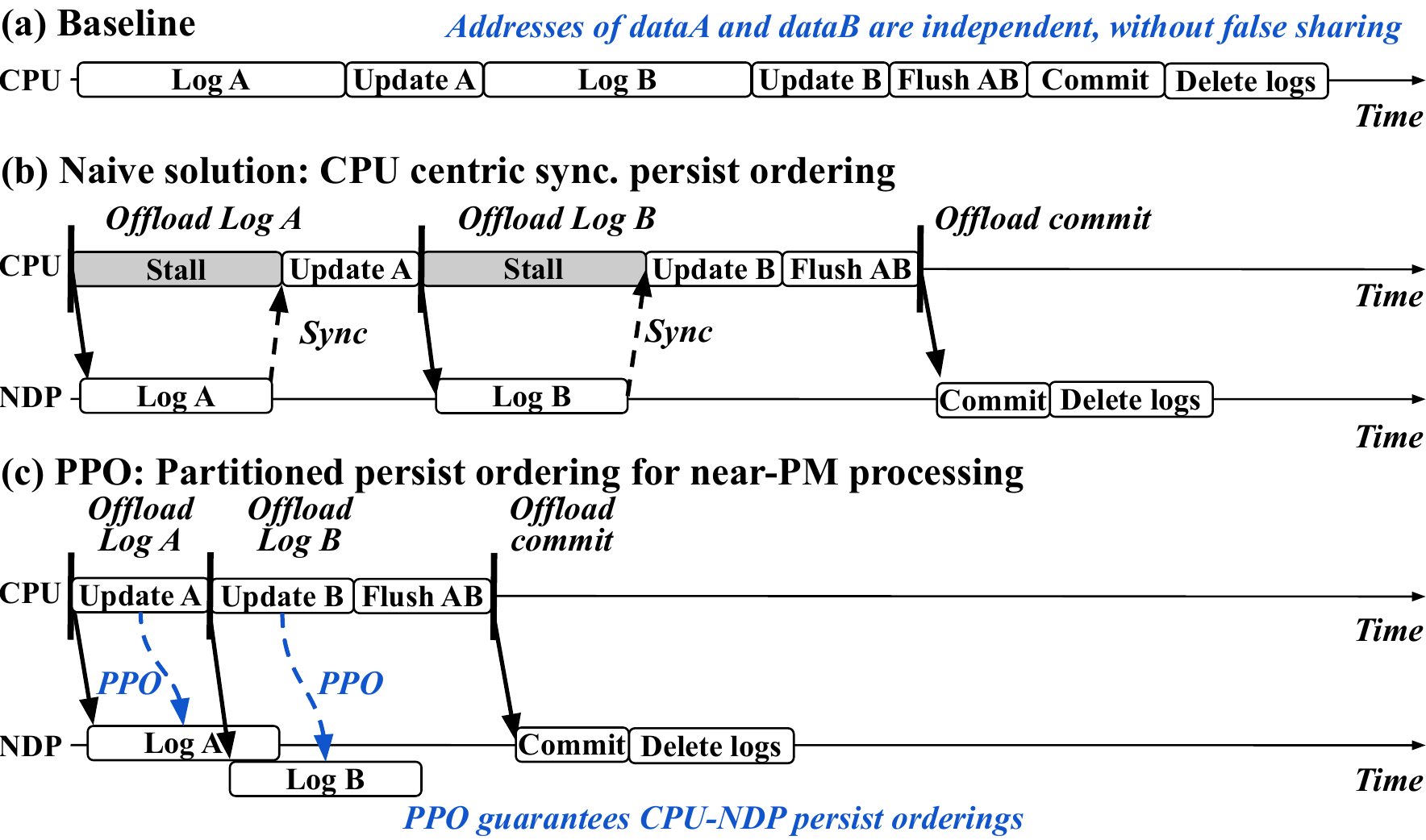}
  \caption{Partitioned execution between CPU and NDP.}\
  \label{fig:undo_decoupled}
\end{figure}

\paragraph*{Persistent Ordering between CPU and NDP device.} 
The major challenge demonstrated in \cref{fig:orderchallenge}a lies in maintaining the persist ordering between the CPU and the NDP. 
To enable a correct persist ordering, a naive solution is to synchronize between the CPU and the NDP device actively. 
As \cref{fig:undo_decoupled}b illustrates, with such a naive solution, execution on the CPU side needs to wait until the execution on NDP has been completed.
Though NDP procedure is faster than the CPU-only baseline (as illustrated in \cref{fig:undo_decoupled}a), frequent synchronization offsets the performance benefits.

However, we observe that maintaining such a strict ordering is not always necessary. 
The partitioned execution on NDP does not always share the memory with the CPU. 
In the example of \cref{fig:neardata}b, the NDP procedure reads from memory but copies and persists it to a \emph{separate memory location} that is only managed by NDP, \ie an undo log.
Therefore, the order of persists to the NDP-managed memory can be relaxed. 
The CPU-side update (\eg ``Update A'' and ``Update B'' in \cref{fig:undo_decoupled}c) needs to persist \emph{after} the associated NDP logging operations (\eg ``Log A'' and ``Log B'' in \cref{fig:undo_decoupled}c). 
While, independent NDP operations, such as logging  different addresses can happen in parallel, without being blocked by the CPU. 
In other crash consistency mechanisms, we observe similar opportunities. 
For example, a page-grained checkpointing operation on NDP only needs to \emph{persist before} any update from the CPU toward the same page, while independent checkpointing operations can persist in parallel as they write to a separate memory location. 
Based on this observation, we see the opportunity to overcome the strict ordering between CPU and NDP units to exploit parallelization.

\begin{figure}
  \centering
 \includegraphics[width=1.0\linewidth]{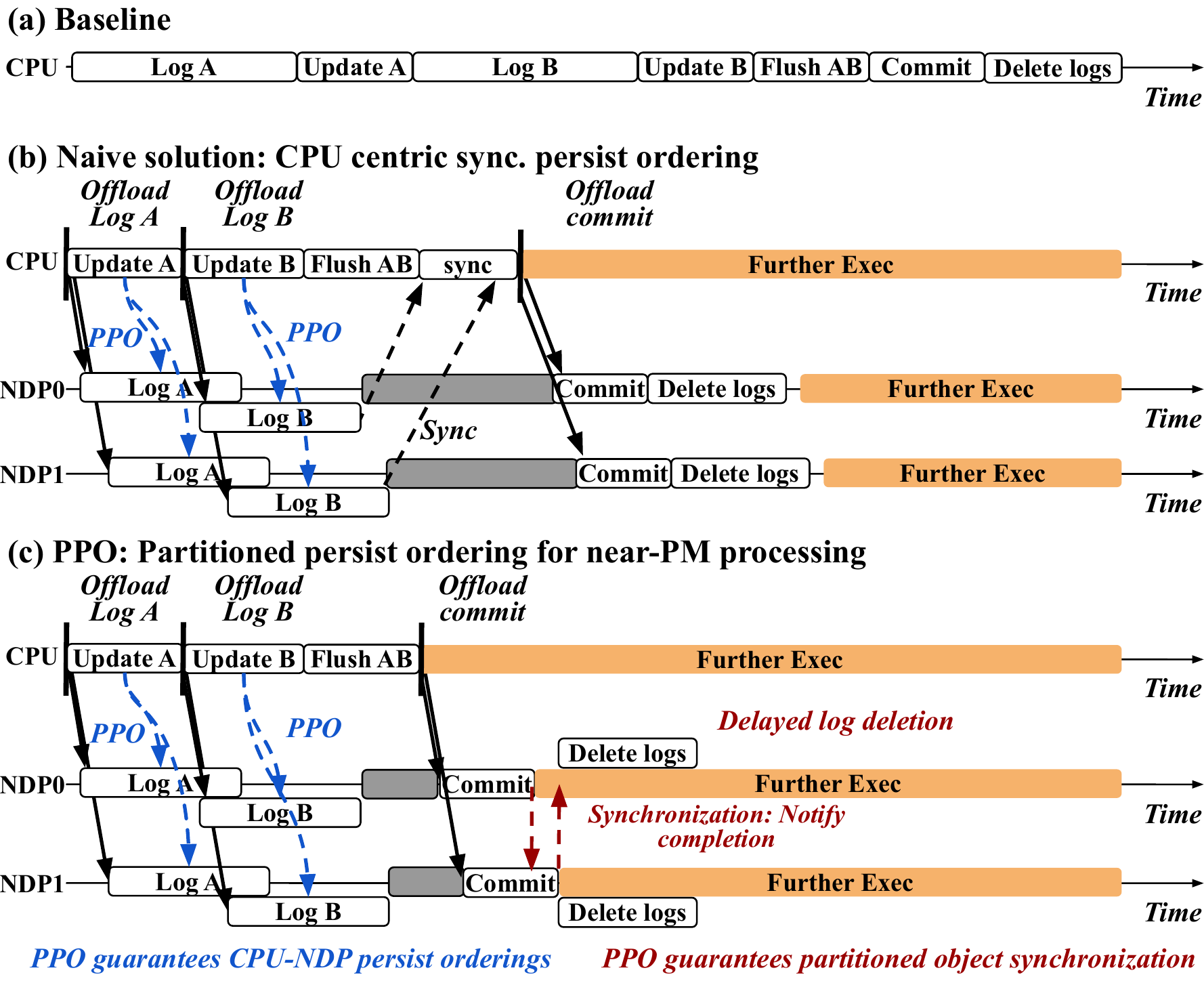}
  \caption{An undo-logging example in multi-device partitioned execution.}\
  \label{fig:delayedlogdelete}
\end{figure}

\paragraph*{Synchronization among multiple NDP devices.}
In addition to CPU and NDP ordering, partitioned execution presents another ordering challenge among multiple devices because a persistent object may span multiple devices \cite{dcpmm_setup}.
The persist ordering among multiple devices is another challenge because the execution is asynchronous among devices and their programs do not necessarily stay at the same pace.
A naive way of maintaining the persist ordering among multiple NDP devices is to actively synchronize all NDP devices after each crash consistency operation to make sure operations on all devices are complete before committing the updates. 
As demonstrated in \cref{fig:delayedlogdelete}b, before updating the data in place (A or B), the CPU stalls before sending a commit operation to the NDP devices, until logging operations on both NDP devices have been completed.
Thus, this naive solution avoids the unrecoverable scenario in \cref{sec:challenges}, as the recovery program either recovers the logged data or keeps the in-place updates. 
Compared to the CPU-centric baseline in \cref{fig:delayedlogdelete}a, even though this naive solution already provides better performance, both the CPU and the NDP devices still stall for synchronization.

\begin{figure}
  \centering
 \includegraphics[width=1.0\linewidth]{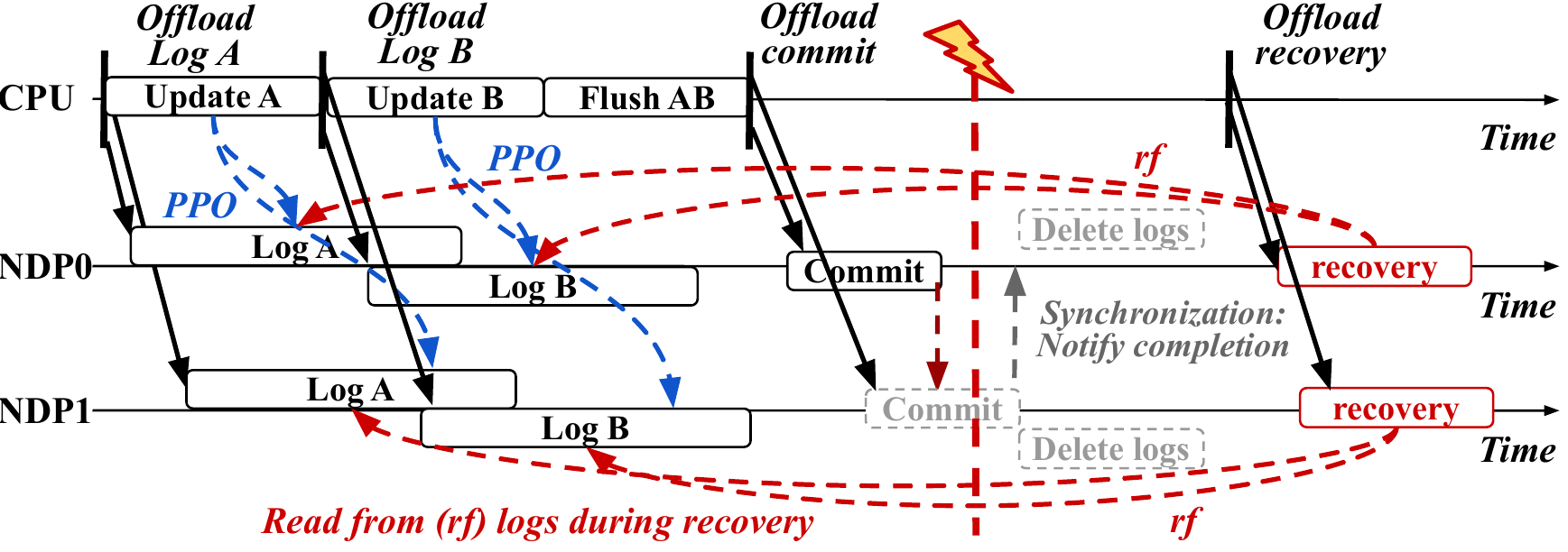}
  \caption{Recovery in multi-device partitioned execution.}\
  \label{fig:resolvechallenge2}
\end{figure}

We further observe that the data required for recovery is only managed by NDP and never exposed to the CPU, unless the recovery procedure happens.
In the example of \cref{fig:delayedlogdelete}c, if we relax the persist ordering between crash consistency operations and the later commit, 
and delay the synchronization among devices, the recovery program can still read from consistent data as long as the data required for recovery is not deleted before the delayed synchronization has completed (\ie ``Delete logs'' for A and B). 
In \cref{fig:resolvechallenge2}, if a failure happens when NDP0 has committed the update but NDP1 has not, the recovery procedure can still read from the consistent copy in the log in both NDP devices, as ``Delete logs'' on both devices only persists after a synchronization.
At the same time, because the synchronization is delayed, it avoids additional stalling on CPU or NDP devices.

\section{Partitioned Persist Ordering}
In this section, we will provide more formal definitions for \pname{} in two scenarios: ordering between CPU and NDP, and among NDP devices.

\subsection{CPU-NDP Ordering} \label{subsec:cpu_ndp_formal}
We first denote an NDP procedure, $N$, that performs a sequence of memory accesses offloaded from CPU. Then, we define basic memory operations to memory address $x$: \\
$\sbullet[0.75]$~$R_{x}$ and $W_{x}$: read and write memory accesses, respectively (from either CPU or NDP).\\
$\sbullet[0.75]$~$R_{x, CPU}$, $W_{x, CPU}$, $R_{x, NDP}$, and $W_{x, NDP}$: CPU read, CPU write, NDP read, and NDP write accesses, respectively.\\
$\sbullet[0.75]$~$R_{x, NDP} \in N$ and  $W_{x, NDP} \in N$ stand for memory read or write issued by NDP procedure $N$ to memory address $x$.\\

Then, we define ordering between memory accesses and NDP procedures:\\
$\sbullet[0.75]$~$\xrightarrow{po}$ denotes program order. \\
$\sbullet[0.75]$~$\xrightarrow{hb}$ denotes happens-before order.\\
$\sbullet[0.75]$~$\leq_{p}$ denotes persist-ordering.

PPO separates out memory addresses that are only managed by NDP procedures, without sharing with CPU. 
NDP accesses to these memory addresses thus do not need to order with memory accesses from CPU.
Maintaining correct execution between CPU and NDP fundamentally depends on two invariants. 
The first invariant concerns the read-write dependency that ensures that execution on CPU or NDP always accesses data in the intended order that was defined by the program. 
The second invariant concerns persistence, as a correct recovery relies on enforcing persist ordering between writes.

\paragraph*{Invariant 1: read-write ordering. }
In memory addresses shared between CPU and NDP, reads and writes issued by an NDP procedure are strictly ordered with the CPU. 
Let us define any read or write from the CPU as $M_{x, CPU}$: $M_{x, CPU}$ $\in \{R_{x, CPU}, W_{x, CPU} \}$, where $x$ is shared between CPU and NDP, i.e., $x \in NDP \bigwedge x \in CPU$.
Likewise, we define $M_{y, NDP}$ that can be either read or write to memory address $y$ that is shared between both CPU and NDP. 
Then, $\forall M_{y, NDP} \in N, M_{x, CPU} \xrightarrow{po} N \Rightarrow M_{x, CPU} \xrightarrow{hb} M_{y, NDP}$, and $N \xrightarrow{po} M_{x, CPU} \Rightarrow M_{y, NDP} \xrightarrow{hb} M_{x, CPU}$.
In essence, within these shared addresses, memory accesses from NDP follow the program order with respect to the CPU.

Whereas, addresses that are only managed by NDP only need to follow the program order within an NDP procedure.
Let $M_{a, NDP} \in \{R_{a, NDP}, W_{a, NDP}\}$, where $a \in NDP \bigwedge a \not\in CPU$
and $M_{b, NDP} \in \{R_{b, NDP}, W_{b, NDP}\}$, where $b \in NDP \bigwedge b \not\in CPU$ be two memory accesses issued by an NDP procedure $N$ to addresses only managed by NDP. 
Then, $M_{a, NDP} \xrightarrow{po} M_{b, NDP} \rightarrow$ $M_{a, NDP} \xrightarrow{hb} M_{b, NDP}$.

\paragraph*{Invariant 2: persistence. }
Like before,  we discuss both NDP-CPU shared memory and NDP-managed memory. 
For memory addresses $x$ and $y$ that are shared between CPU and NDP, the persist ordering follows the program order as well: \\$\forall W_{y, NDP} \in N, W_{x, CPU} \xrightarrow{po} N \Rightarrow W_{x, CPU} \leq_{p} W_{y, NDP}$,\\and $N \xrightarrow{po} W_{x, CPU} \Rightarrow W_{y, NDP} \leq_{p} W_{x, CPU}$.

Writes to NDP-managed memory that is not shared with CPU, say $z$ (e.g., logs, checkpoints, and shadow copies) follow relaxed persist ordering. 
Writes from NDP can delay their persistence, as the CPU cannot access these addresses: \\$\forall W_{z, NDP} \in N,  N \xrightarrow{po} M_{x, CPU} \Rightarrow W_{z, NDP} \not\leq_{p} M_{x, CPU}$.

\subsection{Multiple-Device Synchronization}\label{sec:multiNDPform}
In addition to the definitions in \cref{subsec:cpu_ndp_formal}, we define the following:\\
$\sbullet[0.75]$~$F$: an event of system failure.\\
$\sbullet[0.75]$~$N_{A}$: an NDP procedure executed on NDP device A, which may be interrupted by a failure $F$. Read or write accesses to memory address $x$ that are issued by $N_{A}$ are denoted as $R_{x, NDP}^{A}$ and $W_{x, NDP}^{A}$. \\
$\sbullet[0.75]$~$\mathit{R}_{x, NDP}^A \xrightarrow{rf} \mathit{W}_{x, NDP}^A$: NDP read $R_{x,NDP}^A$ accesses memory address $x$ that was persisted by a write $\mathit{W}_{x, NDP}^A$ before a failure $F$, to recover an interrupted NDP procedure $N_A$. \\
$\sbullet[0.75]$~$S$: a synchronization event that enforces completion and persistence of memory accesses  among all NDP devices. \\

When executing a procedure on a shared object between two NDP devices A and B, if device $A$ synchronizes with $B$ using a synchronization event $S$, then any memory access from the executions of $N_{A}$ and $N_{B}$, may not persist before synchronization is complete, i.e.,
$S \xrightarrow{po} W_{x,NDP_A} \bigwedge S \xrightarrow{po}$ $W_{y,NDP_{B}}$ $\Rightarrow  S \le_{p} W_{x,NDP_{A}} \bigwedge S \le_{p} W_{y,NDP}^B$.

\paragraph*{Invariant 3: Persist before synchronization. }
Before synchronization, any memory access from NDP procedures $N_{A}$ and $N_{B}$ must be persisted, i.e.,
$W_{x,NDP}^{A} \xrightarrow{po} S \bigwedge W_{y,NDP}^{B} \xrightarrow{po}$ $S \Rightarrow W_{x,NDP}^{A} \le_{p} S$ $\bigwedge W_{y,NDP}^{B} \le_{p} S$.
Based on this guarantee, we next discuss the correctness of failure-recovery. 

\paragraph*{Invariant 4: Failure-recovery.}
When the failure happens before synchronization, i.e., $F \xrightarrow{hb} S$, the recovery procedure on each NDP device reads from data that has been persisted before failure for recovery. Say, on NDP device $A$, $R_{x,NDP}^{A} \xrightarrow{rf} W_{x, NDP}^A$, where $W_{x, NDP}^{A}$ has persisted data for recovery. 
As \pname{} enforces persist ordering between writes from NDP and CPU, $R_{x, NDP}^A$ is guaranteed to read consistent data. And the same guarantee applies to device $B$.
When a failure happens after synchronization, i.e., $S \xrightarrow{hb} F$, because all prior memory operations have become persistent, the recovery procedure also reads consistent data.

\section{\xname{} Hardware Design}
\label{mech}

\begin{figure}
  \centering
 \includegraphics[width=\linewidth]{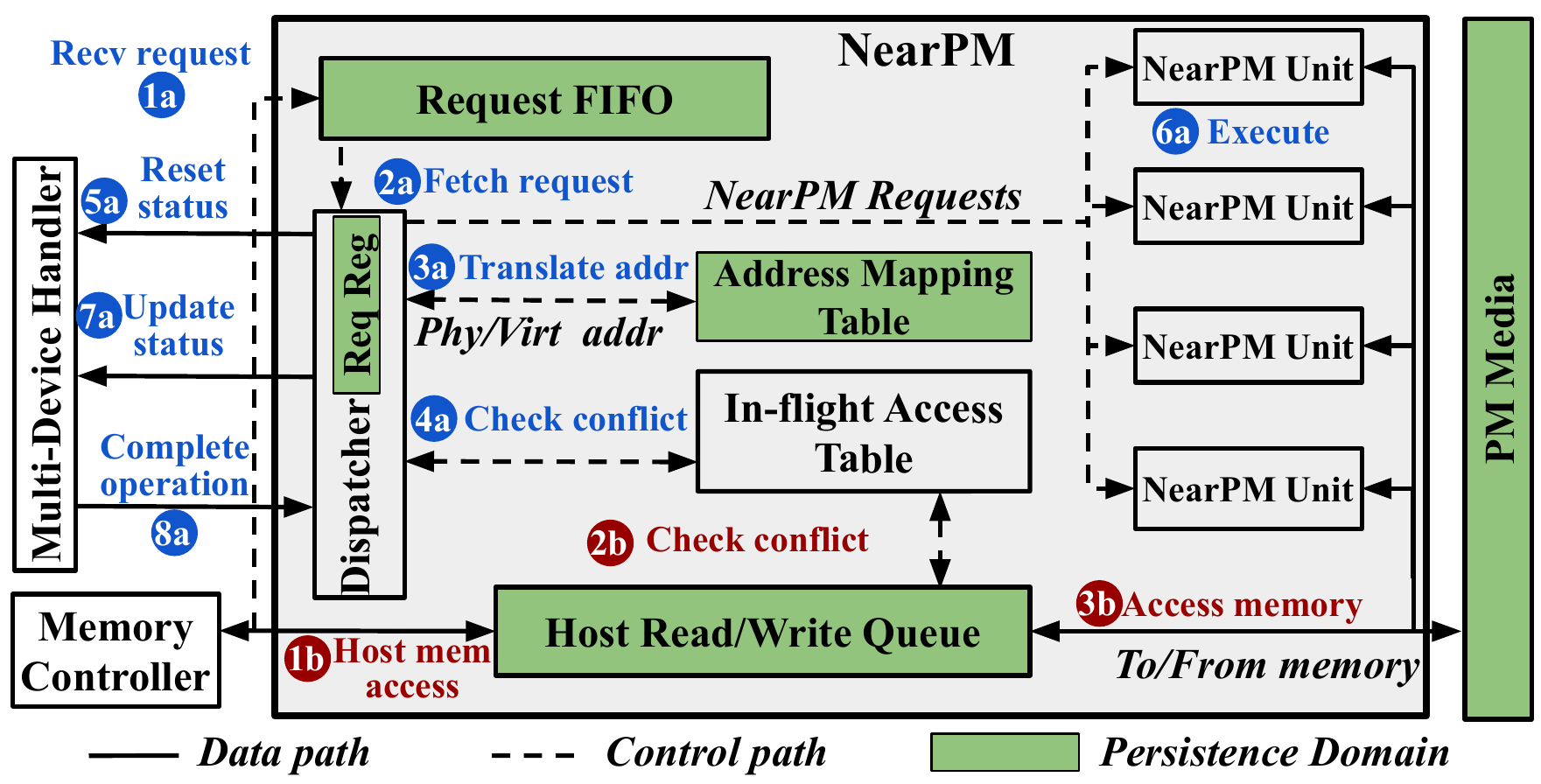}
  \caption{High-level architecture of a \xname{} device.}\
  \label{fig:detailedarch}
  \vspace{-0.15in}
\end{figure}

\subsection{Architecture of \xname{}}
\label{subsec:arch}

\begin{figure}
  \centering
 \includegraphics[width=\linewidth]{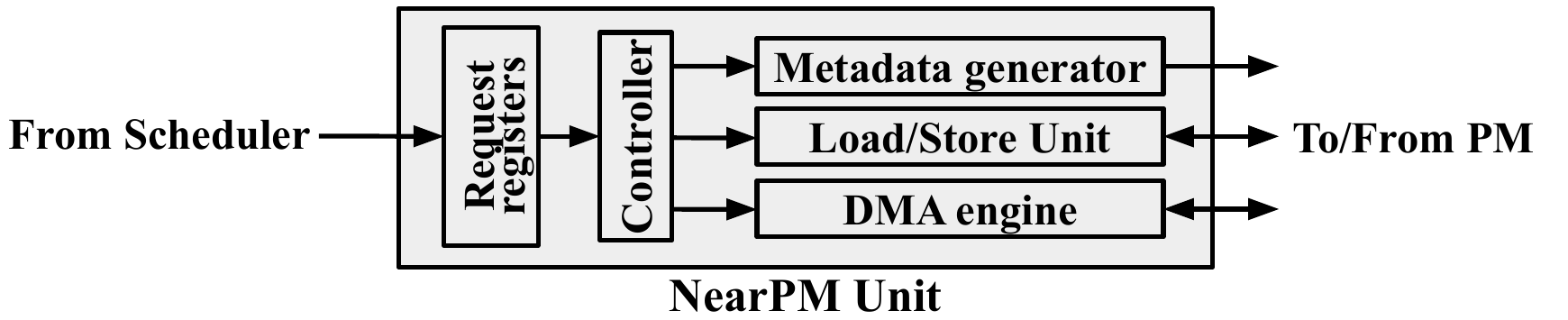}
  \caption{Components in each \xname{} unit.}\
  \label{fig:pmlogunit}
\end{figure}

\xname{} is placed inside the PM controller of the PM device, with direct access to the PM storage medium. 
This enables \xname{} to access PM with higher bandwidth and lower latency than the host processor. 
\xname{} consists of the following major components (Figure \ref{fig:detailedarch}):

\begin{itemize}[leftmargin=*]
    \item \textbf{Host read/write queue} takes regular reads and writes from the host processor and accesses the PM media. 
    \item \textbf{Request FIFO} takes requests issued by the host processor and keeps them until they are executed. 
    \item \textbf{Dispatcher} decodes and issues requests to \textit{\xname{} units} (\ie execution engines). 
    \item \textbf{Address mapping table} converts virtual addresses in the requests to physical addresses, as the parameters \xname{} commands are virtual addresses (details in \cref{subsubsec:address_translation}).
    \item \textbf{In-flight memory access table} keeps track of memory addresses being accessed by the \xname{} units in order to handle accesses with conflicting addresses. In case an operation attempts to access an address that is being written to, the \emph{Dispatcher} stalls this operation and buffers it in the \emph{Host read/write queue}.
    \item \textbf{\xname{} units} are processing engines that manipulate data in PM and are controlled by the \emph{Dispatcher}.
    Each \emph{\xname{} unit} has a request register that stores the request from the \emph{Dispatcher}, a controller which converts requests into control signals, 
    a metadata generator (\eg metadata generation and log deletion), and a load/store unit for fine-grained data movement, and DMA engine for large data movement (\eg data copy), as shown in \cref{fig:pmlogunit}.
    \item \textbf{Multi-device handler} stores the status of other \xname{} devices and coordinates among them. 
    A command execution is complete when all devices have completed execution.
    It keeps track of all \textit{\xname{} units}, issuing a request as soon as one of them is available.
\end{itemize}

\subsection{\xname{} Execution Flow} \label{subsec:exec_flow}
In this section, we further introduce the execution flow of \xname{} that handles program-offloaded crash consistency operations (\ie \xname{} requests) and services the regular memory accesses from the host processor.

\paragraph*{\xname{} request execution.}
\cref{fig:detailedarch} shows the workflow (steps in blue).  
A \xname{} request first enters the \emph{Request FIFO} (step \circledtext{1a}) and then gets decoded by the \emph{Dispatcher} (step \circledtext{2a}).
During decoding, the \emph{Dispatcher} translates  request operands from virtual to physical address through an \emph{Address Mapping Table} (step \circledtext{3a}).
After translation, the \emph{Dispatcher} checks the request's physical address (step \circledtext{4a})---requests without address conflicts are immediately issued, but stall until the completion of the other conflicting request/access (details in \cref{dephandle}).
Next, \xname resets the status bit in \emph{Multi-device handler} (step \circledtext{5a}).
Then, \xname{} unit receives the request and starts the execution immediately (step \circledtext{6a}).
Upon completion, \xname{} notifies the \emph{Multi-device handler} to update the status bit both locally and in other \xname{} devices (step \circledtext{7a}).
When all \xname{} devices have completed execution, the \emph{Multi-device handler} notifies the \emph{Dispatcher} (step \circledtext{8a}) to assign new commands to the \xname{} unit. 

\paragraph*{Host memory access.}
Figure~\ref{fig:detailedarch} (steps in red) describes the execution for CPU's memory accesses.
CPU's memory accesses enter the host read/write queue (step \circledtext{1b}). 
Like before, the \emph{Dispatcher} also checks the CPU's accesses for address conflicts before dispatching (step \circledtext{2b}). 
It issues memory access immediately if there is no conflict from the \emph{In-flight Access Table} (step \circledtext{3b}), Otherwise, it buffers CPU's access until the other access has completed.

\subsection{Correctness Guarantees}

\subsubsection{CPU-NDP Ordering.}
\label{dephandle}

\xname{} implementation follows \pname{}.
We first discuss the implementation that ensures CPU-NDP ordering. 

\begin{figure}
\vspace{7pt}
  \centering
 \includegraphics[width=\linewidth]{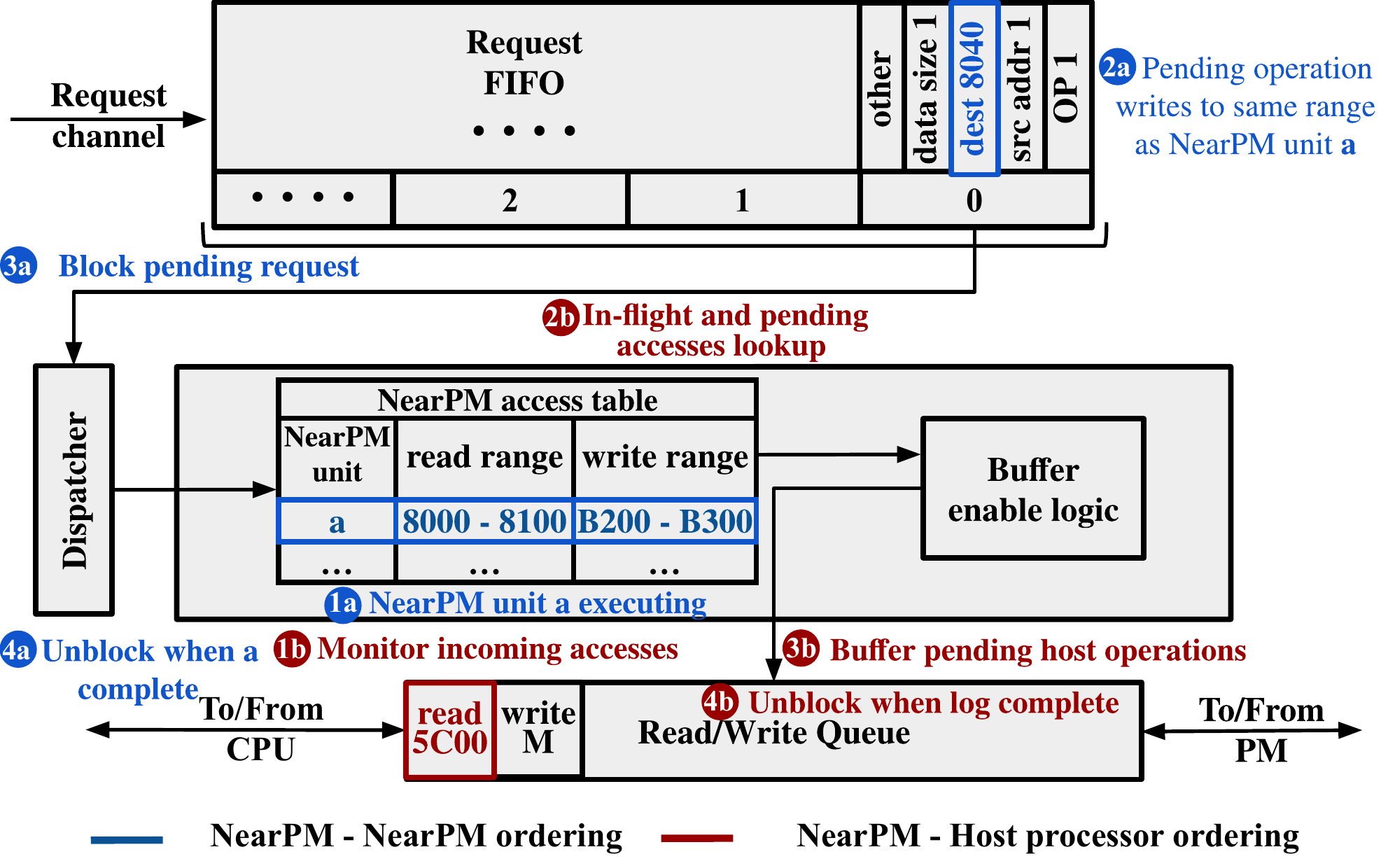}
  \caption{Ordering handling in a single \xname{} device.}
  \label{fig:depend}
\end{figure}

Invariant 1 is ensured by the \emph{Dispatcher} (\cref{subsec:arch}).
When \xname{} dispatches a request to a \xname{} unit for execution, it updates addresses in the \textit{\xname{} access table} (Figure \ref{fig:depend})
When the CPU accesses PM (Figure \ref{fig:depend}  step \circledtext{1b}),  \xname{} checks if there are ordering dependencies between inflight \xname{} execution and incoming CPU memory accesses (step \circledtext{2b}).
If a conflict is detected, \xname{} buffers incoming memory access from the host (step \circledtext{3b}) until the conflicting access is completed by the \xname{} units (step~\circledtext{4b}).
To avoid ordering invariants violations between NDP procedures in a single \xname{} device, the \emph{Dispatcher} checks the lookup table for conflicts between memory ranges accessed by the pending and in-flight requests (step \circledtext{2a}). If there is a conflict, the \emph{Dispatcher} delays the issue of pending requests (step \circledtext{3a}) until the conflicting access has completed (step \circledtext{4a}). 

Invariant 2 is ensured by writing back all updates to PM on the CPU side before invoking an NDP procedure.
As there is no write caching in the NDP device, as soon as NDP issues a write access to PM, it enters the persistence domain. 

\subsubsection{NDP-NDP Ordering}
\label{sec:imp_multidev}

\begin{figure}
  \centering
  \includegraphics[width=1.0\linewidth]{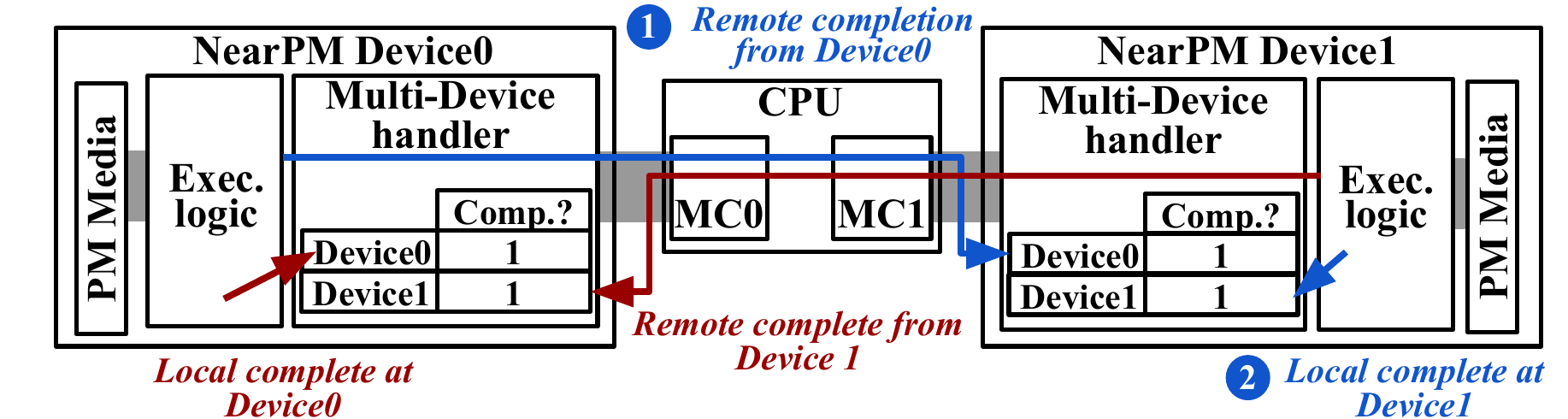}
  \caption{Hardware for cross-device synchronization.}\
  \label{fig:command_sync_hw}
\end{figure}

\pname{} enables delayed synchronization because writes to data required for recovery do not need to complete immediately.
Therefore, synchronization is not on the critical path. 
We take an approach described in \cref{fig:command_sync_hw} to coordinate the completion of requests among devices.
When the program issues a \xname{} request that operates on a persistent object spanning multiple devices, similar to issuing memory requests in interleaved memory modules, the memory controller sends the \xname{} request to all interleaved \xname{} devices according to their address ranges. 
Each \xname{} device has a \emph{Multi-device handler} that keeps track of the status of each command in local \xname{} execution logic as well as in other \xname{} devices.
After \xname{} starts execution, it waits for the completion status from other \xname{} devices (step \circledtext{1}) and its local execution (step \circledtext{2}). 
Finally, \xname{} removes or resets the data required for recovery, which is not on the critical path of execution.
In this way, the delayed synchronization mitigates the performance overheads.

\begin{figure}
  \centering
  \includegraphics[width=1.0\linewidth]{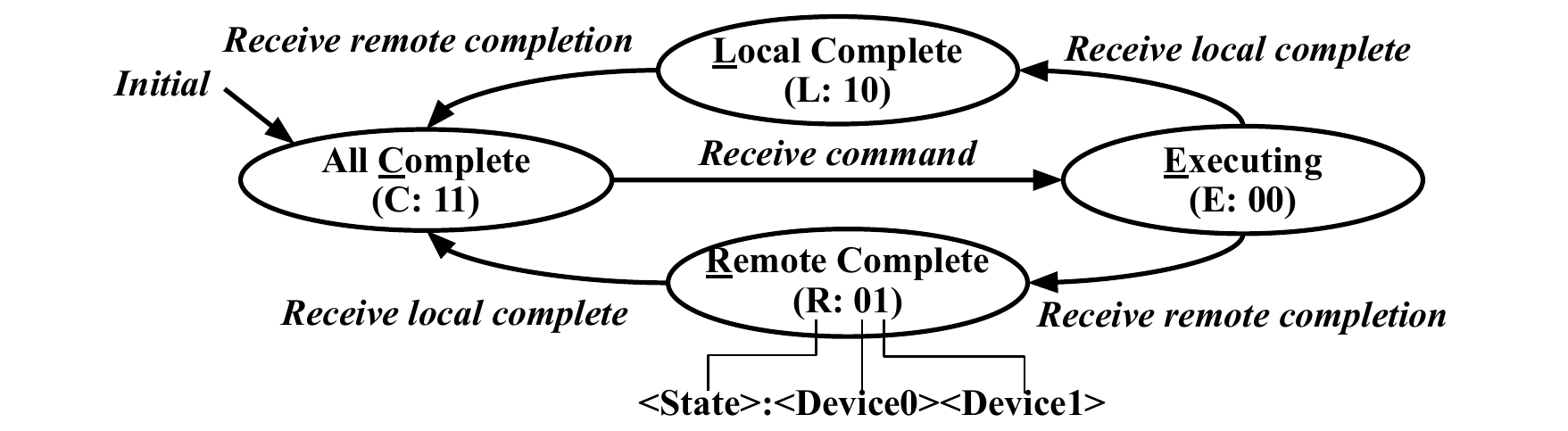}
  \caption{Synchronization state machine of partitioned execution on two devices.}\
  \label{fig:command_sync_state}
\end{figure}

\pname{} maintains a state machine to keep track of the synchronization status during partitioned execution to meet invariant 3.
\cref{fig:command_sync_state} shows a state machine for a two-device setup.
The state machine starts from the \emph{All Complete (C)} state until a command that was duplicated to execute in two devices is received. 
Then the state machine changes its state to \emph{Executing (E)} and keeps monitoring for \emph{Receive local complete} or \emph{Receive remote complete} signals from local execution or remote execution. 
After receiving the command complete signals from all devices, it will return back to the \emph{All Complete} state.
When both devices reach \emph{All Complete} state, writes before this point have become persistent.

\subsubsection{Recovery} \label{subsec:recovery}

Failure-recovery is another aspect of correctness.
To satisfy Invariant 4, the NDP system keeps in-flight operations in a persistence domain and restores them after failure.

\paragraph*{Persistence domain.}
PM hardware systems employ extended persistence domains (e.g., ADR \cite{ADR} and eADR \cite{rudolf_persist_cache}) that include not only the PM devices but also buffers/caches in the processor.
As \xname{} executes the crash consistency operations in the PM module and services regular memory accesses from the host processor, these operations and memory access requests should also be placed in the persistence domain, in case they are not completed before failure.
\cref{fig:detailedarch} marks the hardware components of \xname{} within the persistence domain in green: Request FIFO (2 kB), Address Look-Up Table in the Address translator (432 Bytes), In-flight request registers (256 Bytes) in the Dispatcher, and Host Read/Write Queue (4 kB).
Those structures have a total capacity of 7 kB, much less than the buffers (tens of kBs) in existing Optane PM modules \cite{wang20_micro}). 
Thus, it is practical to use residual capacitors similar to existing Optane PM to write back structures in the persistence domain to a reserved PM location upon failure.

\paragraph*{Recovery procedure.}
After the system is up again, the hardware of a \xname{} device ensures that the results of in-flight \xname{} requests and pending host memory accesses in the persistent domain are visible to the recovery program. In a case where there are multiple \xname{} devices in the system, the recovery program needs to determine the progress made by each device prior to failure---the latest synchronization point that all \xname{} devices reach before failure happens.
The recovery procedure of \xname{} hardware includes two steps: 
(1) \xname{} loads the data from the reserved PM region back to the structures in the persistence domain.
(2) \xname{} replays the in-flight \xname{} requests and host memory accesses until it reaches the latest synchronization point. Thus, the results of all in-flight operations prior to the synchronization point are visible in memory. 

\subsection{Address Translation}\label{subsubsec:address_translation}

Address translation has always been a challenge in NDP systems \cite{gao2015_PACT,xi2015beyond, boroumand2018asplos,fernandez2020_ICCD,gao2015_PACT,gao2016_HPCA,hsieh2016_sigarch,hsieh2016_ICCD,kim2016_sigarch,kim2018_BMC,mutlu2020modern,singh2020_FPL,singh2019_DAC,zhan16_micro,kim2018_BMC} as structures such as TLB are in the host processor. 
Fortunately, PM libraries (e.g.\cite{pmdk}) usually allocate PM as pools and a memory access to the pool manifests as a base address plus an offset within the pool. 
Prior works have shown that as long as a pool's base address is translated, it is straightforward to also translate other memory addresses in the same pool using the offset value~\cite{wang17_micro, wang18_isca, ye2021supporting}.
Therefore, \xname{} keeps the translation of the base address for each pool and performs address translation without going through the CPU.

\begin{figure}
\vspace{7pt}
  \centering
 \includegraphics[width=\linewidth]{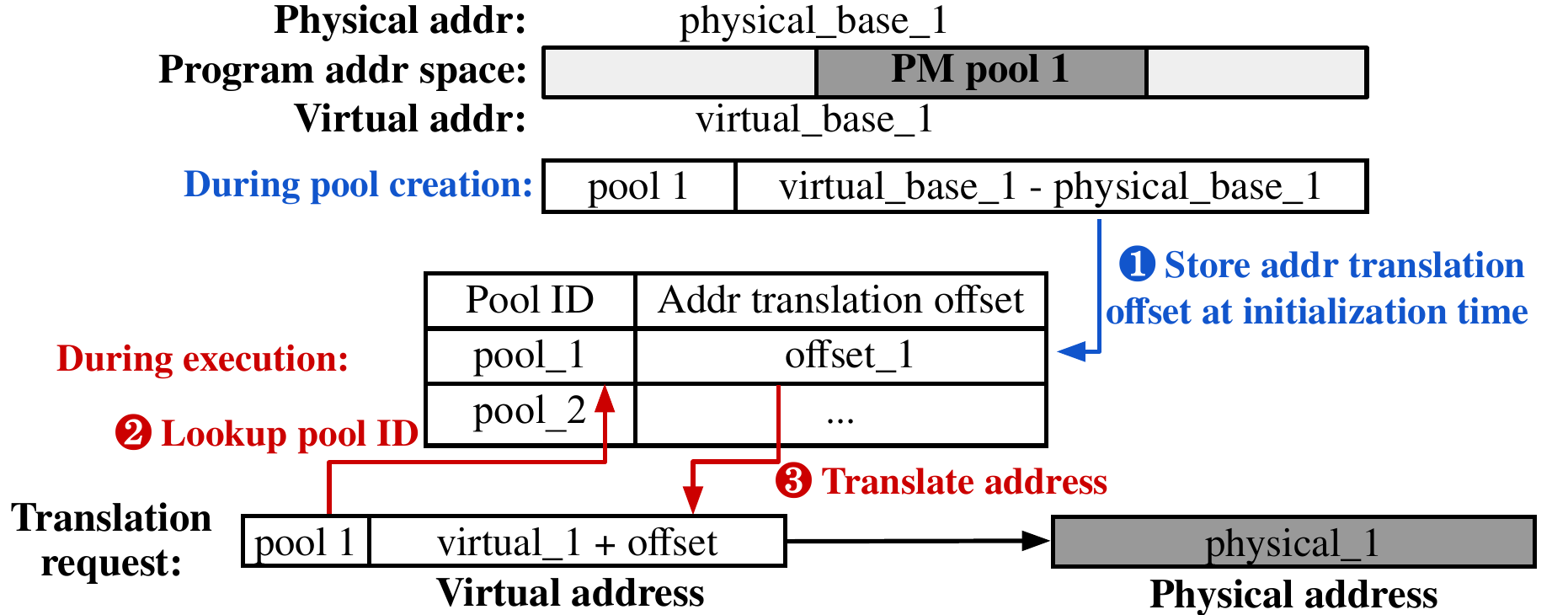}
  \caption{Address translation in \xname{}.}\
  \label{fig:addrtranslate}
  \vspace{-0.15in}
\end{figure}

\sloppy{
Figure {\ref{fig:addrtranslate}} shows the address translation procedure in \xname{}. 
When the program creates a PM pool, \xname{} first computes the offset between the virtual and physical addresses.
The offset is then stored in the \emph{Address Mapping Table} indexed by the pool ID (step {\circledtext{1}}).
Because the addresses encoded in the command are from the virtual address space, to execute them, \xname{} looks up the pool ID of the incoming request (step {\circledtext{2}}) and translates its virtual address to the physical address, by adding the offset to the incoming virtual address (step {\circledtext{3}}).
When accommodating multi-threaded applications, in addition to the pool ID, thread ID is also used for indexing address translation offset.
}

\paragraph*{Context switch handling.}
\xname{} keeps the base address mapping for each PM pool. 
As each pool ID is unique in the system, even across a context switch, the pool-ID-indexed translation mapping still remains valid.

\paragraph*{Multi-device support.}
A PM pool can span across multiple interleaved \xname{} devices, where certain bits in the virtual address identifies which \xname{} device the data locates. 
Based on these bits, each device contains a virtual-physical mapping for the base address that is mapped to its local device. 
Thus, the translation mechanism that relies on the base address of the pool still applies to  multi-device scenarios.

\begin{figure}
  \centering
 \includegraphics[width=\linewidth]{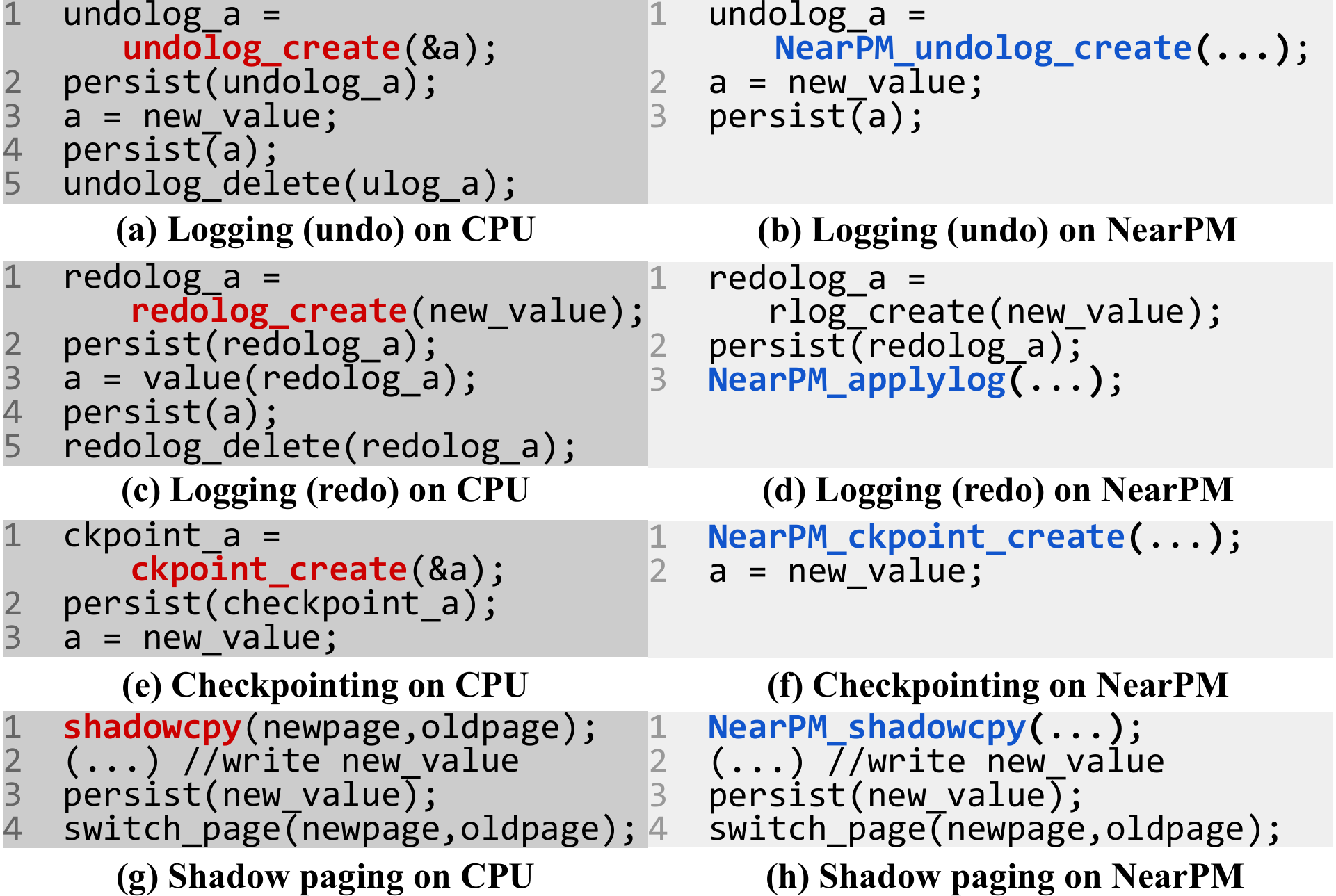}
  \caption{Examples that demonstrate the use of \xname{}'s software interface.}
  \label{fig:codeexample}
\end{figure}

\begin{table*}[t]
    \aboverulesep = 0.22em
    \belowrulesep = 0.22em
    
    \caption{\xname{} software interface.}
    \label{tab:apibreakdown1}
    \setlength{\tabcolsep}{3pt}
    \small
        \centering
        \footnotesize

        \begin{tabular}{lll}
            \toprule
            \textbf{\xname{} primitives} &         
            \textbf{Arguments}  & \textbf{Description}\\
            \midrule
            NearPM\_undolg\_create & pool\_id, thread\_id, old\_data\_ptr, size & Undo-logging: generate metadata and copy old data to an undo log \\ 
            \midrule
            NearPM\_applylog & pool\_id, thread\_id, redolog\_ptr, size  & Redo-logging: apply a redo log by copying data to the original location \\ 
            \midrule
            NearPM\_commit\_log & pool\_id, thread\_id & Undo/redo-logging: delete and commit multiple logs in a PMDK transaction \\
            \midrule
           NearPM\_ckpoint\_create & pool\_id, thread\_id, old\_data\_ptr, size & Checkpointing: generate metadata and copy existing data to a checkpoint before update   \\
            \midrule
           NearPM\_shadowcpy  & pool\_id, thread\_id, page\_ptr, size & Shadow-copy: copy an existing page before update  \\
              \midrule
           NearPM\_init\_device &  device\_path & Device initialization: memory-maps  \xname{}'s command interface\\
            \bottomrule
        \end{tabular}
\end{table*}

\section{\xname{} Software Design}

\xname{} is a software-hardware co-design, including a software interface for developing programs and communication between the program and \xname{}.

\subsection{Software Interface}
\label{sec:swinterface}
For the API to communicate with the \xname{} hardware, there is a dedicated control path (\cref{fig:detailedarch}).
This control path is memory-mapped using a separate address space from that of the PM. 
At the application level, the program needs to provide the device path of \xname{} to \texttt{NearPM\_init\_device} which memory-maps this command path.

The PM program running on the host processor uses a software API to issue commands.
\cref{tab:apibreakdown1} lists the primitive functions and their parameters, which can be directly called in PM programs or libraries.
In our evaluation, we implemented the APIs in the PMDK \cite{pmdk} library. 
The API is agnostic whether or not an operation is single or multi-device. 
The address range of the command operand is monitored by the memory controller; a command is duplicated if the operand object is shared among multiple devices.
\cref{fig:codeexample} demonstrates code examples that show the use of these calls.
Each primitive function in the API corresponds to the crash consistency mechanisms in \cref{tab:commonops}.
Besides the listed primitive functions, our prototyping system can also be extended to test and develop other crash consistency mechanisms.

\subsection{Recovery}

When recovering back from a system failure the software is responsible for initializing the hardware recovery procedure. 
The recovery program sends the command for system-wide recovery and the \xname{} devices will individually run their own recovery procedures following the hardware steps discussed in \cref{subsec:recovery}
After completion, the PM program can start executing from the last consistent program point.

\section{\xname{} Implementation}
\label{sec:implementation}

We use the Xilinx Virtex UltraScale+ VCU118 evaluation platform \cite{vcu118} to implement \xname{}. The development board is attached to a PCIe $3.0\times8$ slot, with a bandwidth of 8 GB/s. 
We use 2 GB of the onboard DRAM to emulate PM on a \xname{} device. 
In our evaluation, the access latency to the emulated PM is 436 ns, similar to real evaluations on Intel's Optane DCPMM \cite{izraelevitz19_dcpmm}.
\xname{} connects to the CPU's memory controller via the PCIe bus.
Constrained by the FPGA platform, we implement two \xname{} devices on the same FPGA, with each device having four \xname Units running at 300MHz.
The emulated \xname{} devices on the FPGA are mapped to a contiguous memory region. 
\xname{} can only support interleaving which will result in a contiguous block in a given device. 
Scatter-gather operations are not supported.
Each \xname{} device contains four \xname{} units, connected through an internal AXI bus of 4 GB/s bandwidth.

The DRAM on the FPGA board is mapped to the CPU's physical memory space in the write-back cacheable mode and is directly accessible through load-store instructions.
However, the Linux kernel maps FPGA's memory as non-cacheable by default. Thus, we manipulate the memory type range register (MTRR) at the boot time to enable writeback caching. 
Thus, our implementation is a software-implemented coherent memory. 
In the upcoming CXL \cite{cxl} systems, we expect even better performance.

\begin{figure*}
    \centering
    \begin{subfigure}[t]{0.32\linewidth}
        \includegraphics[height=1.26in]{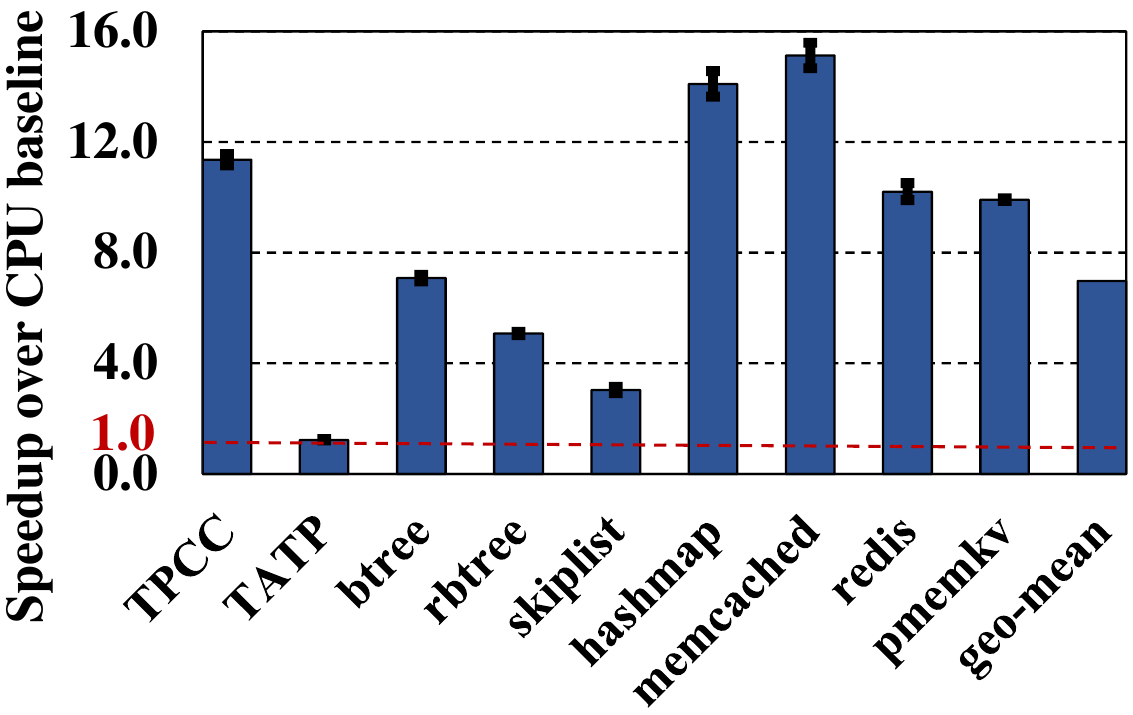}
        \caption{}
    \end{subfigure}
     \hspace{-0.03\linewidth}
    \begin{subfigure}[t]{0.3\linewidth}
        \centering
        \includegraphics[height=1.25in]{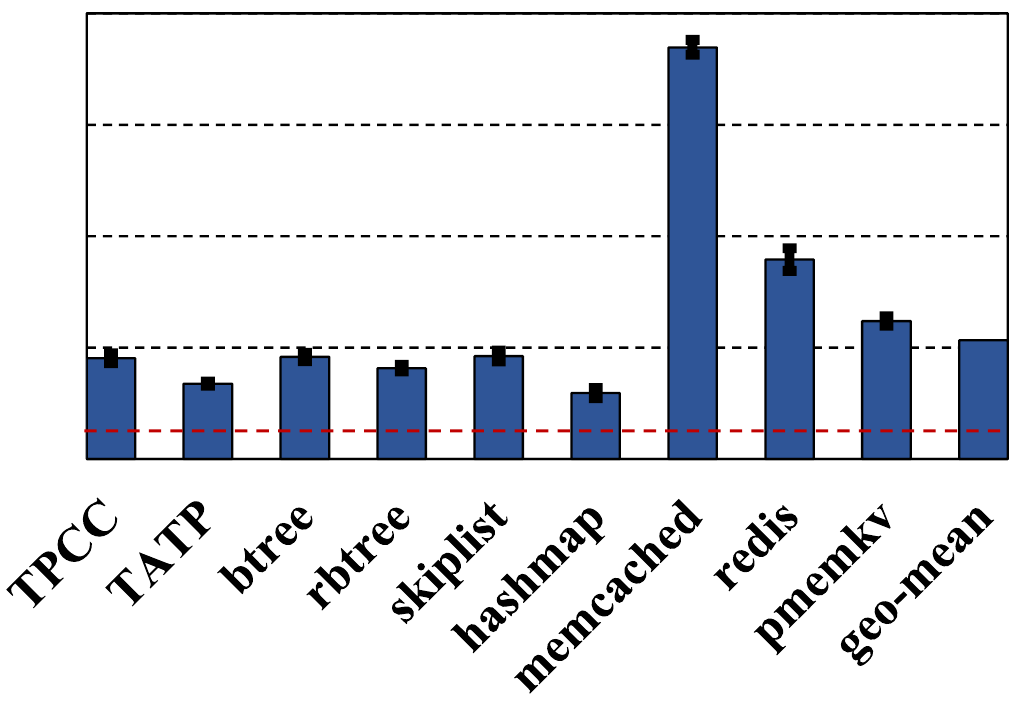}
        \caption{}\
    \end{subfigure}
    \begin{subfigure}[t]{0.29\linewidth}
        \centering
        \includegraphics[height=1.25in]{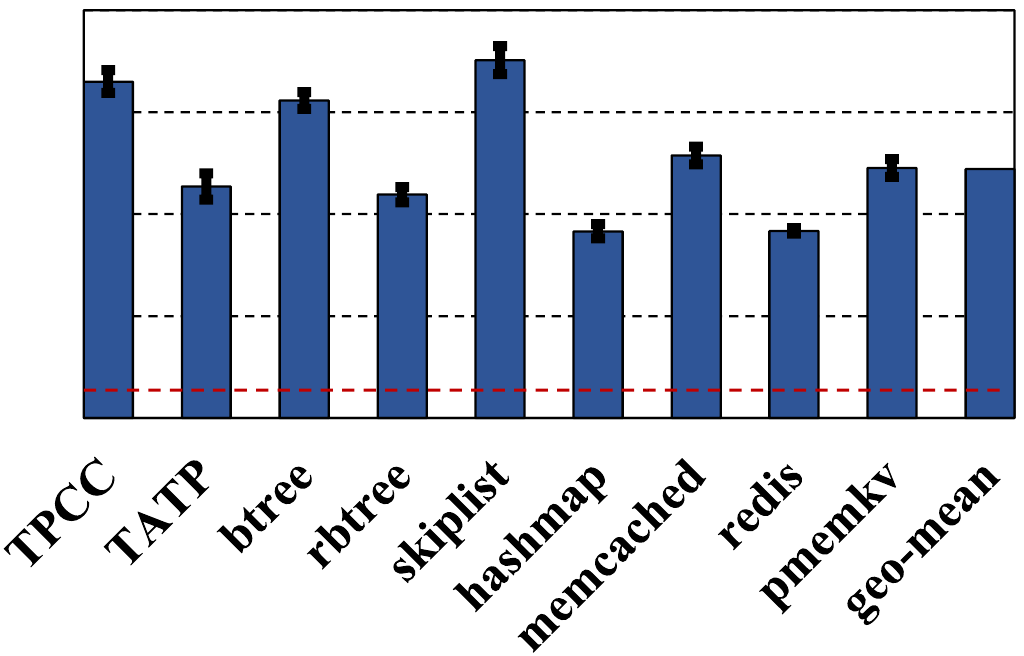}
        \caption{}\
    \end{subfigure}
    \vspace{-0.2in}
    \caption{Speedup in code regions for crash consistency in (a) logging, (b) checkpointing, and (c) shadow paging.}
    \label{fig:crashp_kernel}
    \end{figure*}

\begin{figure*}
    \centering
    \begin{subfigure}[t]{0.32\linewidth}
        \centering
        \includegraphics[height=1.15in]{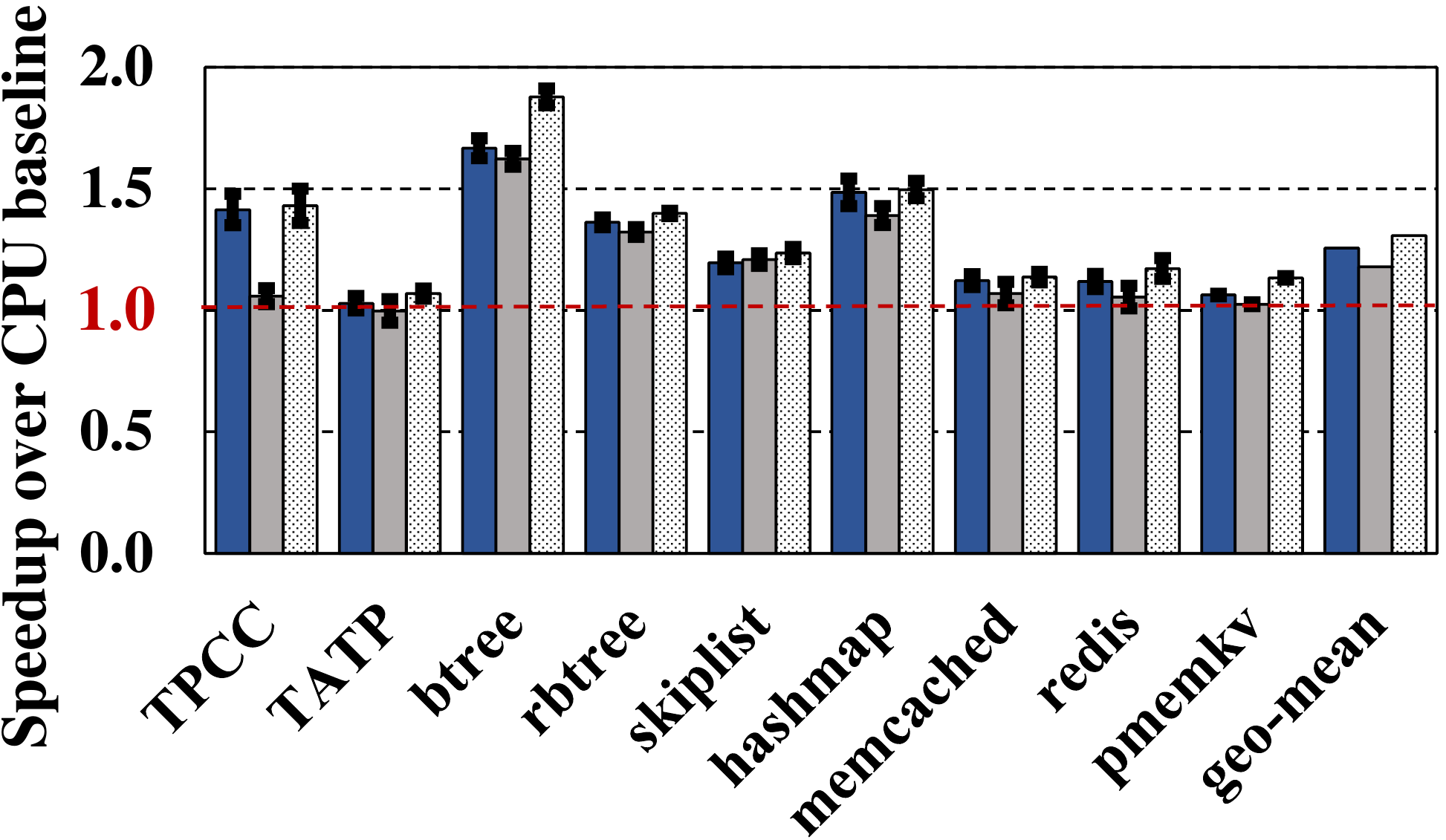}
        \caption{}
    \end{subfigure}
    \begin{subfigure}[t]{0.3\linewidth}
        \centering
        \includegraphics[height=1.25in]{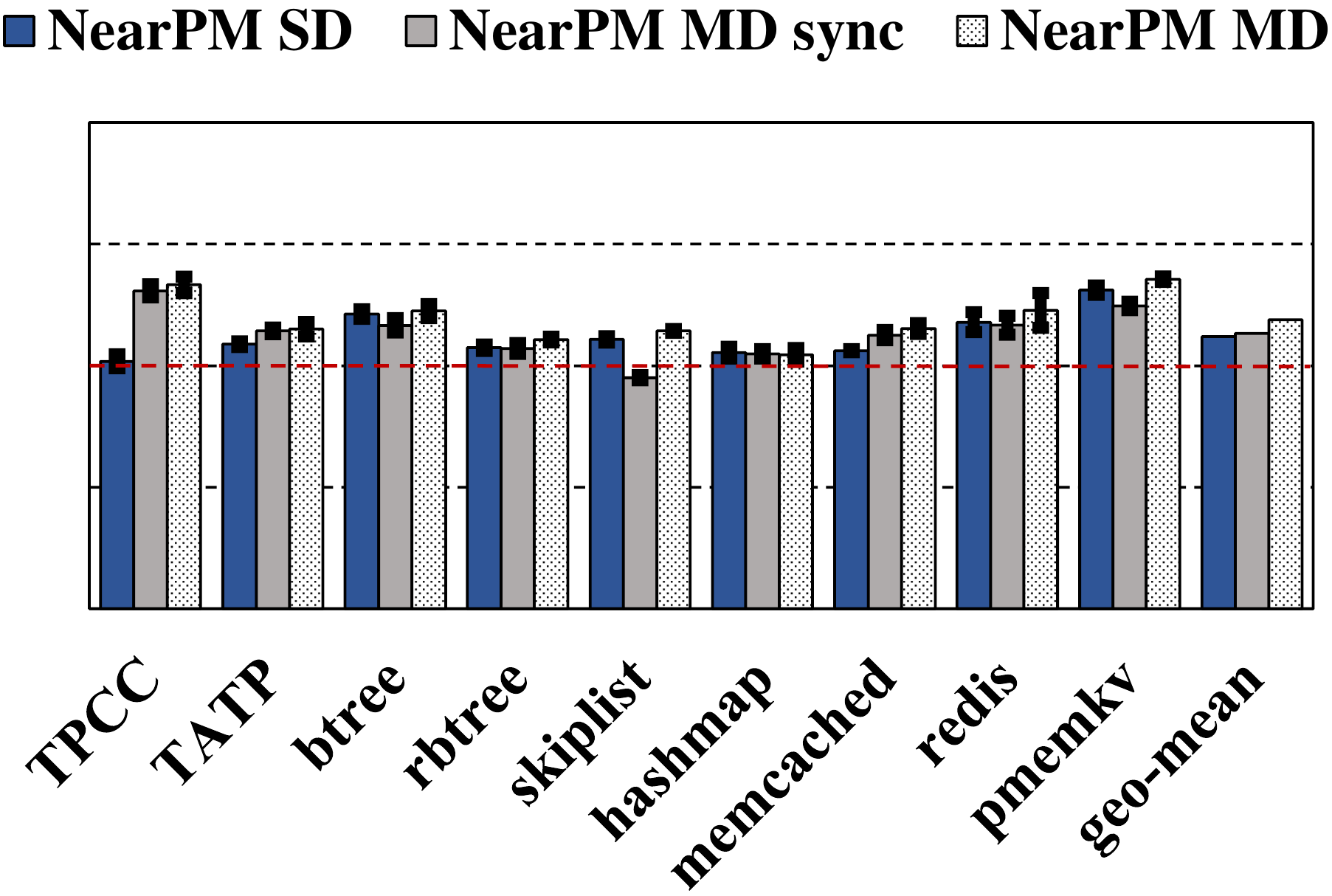}
        \caption{}\
    \end{subfigure}
    \begin{subfigure}[t]{0.29\linewidth}
        \centering
        \includegraphics[height=1.1in]{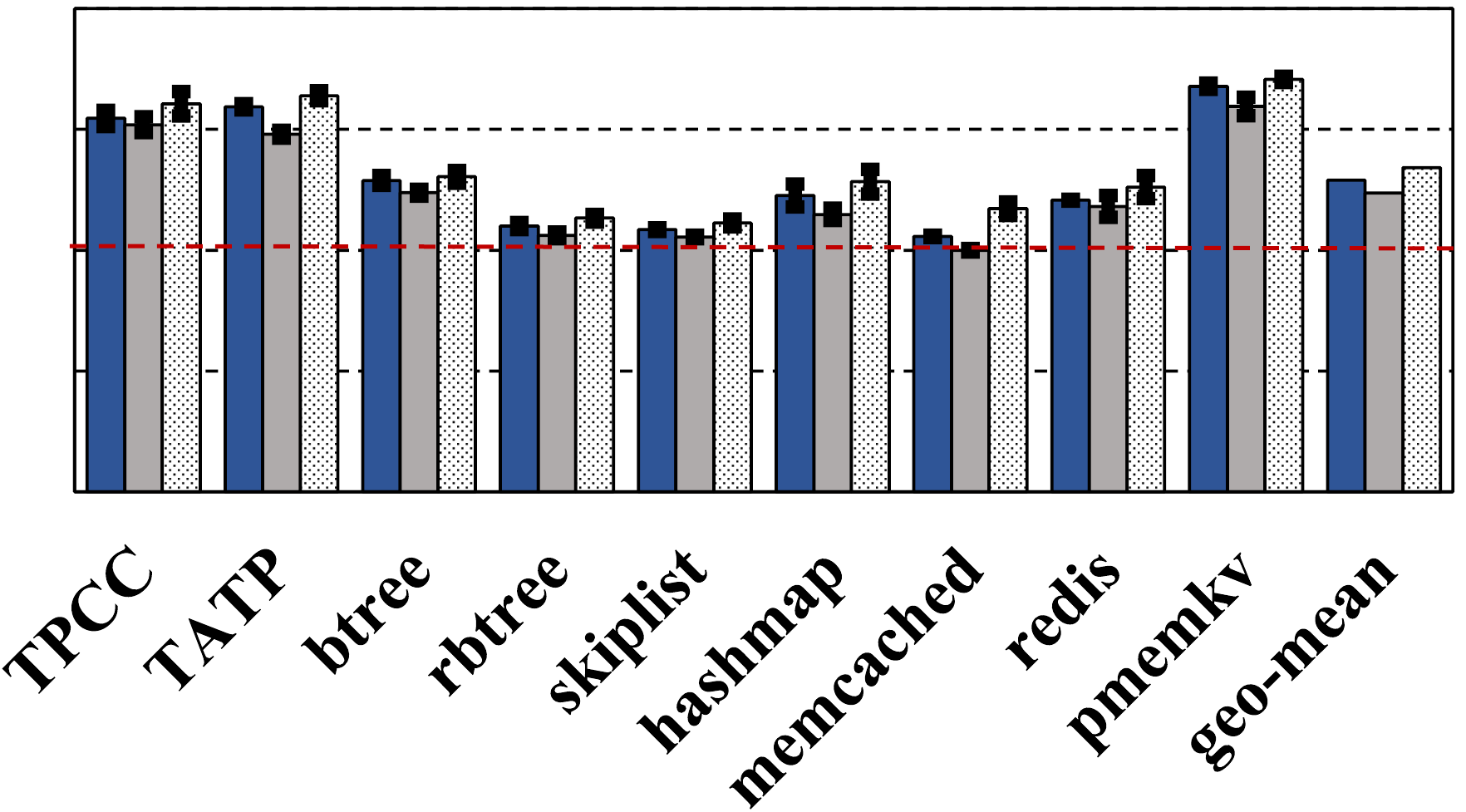}
        \caption{}\
    \end{subfigure}
    \vspace{-0.2in}
    \caption{End-to-end speedup in (a) logging, (b) checkpointing, and (c) shadow paging.}
    \label{fig:crashp}
\end{figure*}

\begin{table}[t]
\aboverulesep = 0.22em
\belowrulesep = 0.22em
\caption{System for evaluation.}
\vspace{-0.05in}
\label{tab:system_config}
\setlength{\tabcolsep}{3pt}
    \centering
    \footnotesize
    \begin{tabular}{l l}
        \toprule
        \multicolumn{2}{c}{\textbf{System Configuration}} \\
        \midrule
        CPU &  AMD Zen 2, 2 GHz, 8 cores\\
        DRAM & 4$\times$16 GB DDR4 \\
        FPGA & Xilinx UltraScale+ VCU118 (Section~\ref{sec:implementation})\\
        PM & 2 GB, Emulated with on-FPGA DRAM \\
        \xname{} & 4 \xname{} units, 32 entry request FIFO \\
        \midrule
        \multicolumn{2}{c}{\textbf{Software System}} \\
        \midrule
        OS & Ubuntu 20.04, Linux kernel v5.3.0 \\
        Environment & gcc/g++-9.2, PMDK-1.9 \\
        \bottomrule
    \end{tabular}
\end{table}

\begin{table}
\aboverulesep = 0.22em
\belowrulesep = 0.22em
\caption{Workloads for evaluation.} \label{tab:workloads}
\setlength{\tabcolsep}{1pt}
    \centering
    \footnotesize    
    \begin{tabular}{L{3.3cm}L{4.5cm}}
        \toprule
        \textbf{Workload} & \textbf{Input} \\
        \midrule
        TPCC, TATP \cite{gogte18_pldi} & Process TPCC/TATP transactions    \\
        \midrule
        btree, rbtree, skiplist, hashmap \cite{pmdk} &  Insert random key-values (value size is 64 B)   \\
        \midrule
        memcached  \cite{pmem-memcached}   & 100\% write requst from  YCSB \cite{YCSB}  \\
        \midrule
        redis \cite{pmem_redis} & 100\% write requst from YCSB \cite{YCSB}   \\
        \midrule
        pmemkv  \cite{pmemkv} & Input from PmemKV-bench \cite{pmemkv-bench}\\
        \bottomrule
    \end{tabular}
\end{table}

\section{Evaluation} \label{sec:eval}

\subsection{Methodology} \label{subsec:methodology}

\paragraph*{System configuration.}
We evaluate \pname{} on the prototype of \xname{} (implementation in \cref{sec:implementation}) in a testbed described with the system configurations in \cref{tab:system_config}.

\paragraph*{Workloads.}
Table~\ref{tab:workloads} lists the workloads and their inputs.
TPCC and TATP are PM transactions from a prior work \cite{gogte18_pldi};
btree, rbtree, skiplist, and hashmap are key-value stores from PMDK \cite{pmdk} library;
Redis and Memcached are real-world workloads. 
PmemKV \cite{pmemkv} is a key-value store that uses a B+ tree as the backend.
For each workload, we evaluate three crash consistency implementations:
\begin{itemize}[leftmargin=10pt]
\item {\textbf{Logging:}} The performance of each program's original crash consistency support based on undo/redo logging.
\item {\textbf{Checkpointing:}} The performance of a modified crash consistency support based on checkpointing.
\item {\textbf{Shadow paging:}} The performance of a modified crash consistency support based on shadow paging.
\end{itemize}
Note that both checkpointing and shadow paging operate at 4 kB page granularity.

\paragraph*{Comparison points.}
We evaluate  four configurations, where all experiments are evaluated 10 times. The error bars (standard deviation) are included in the figures.

\begin{itemize}[leftmargin=10pt]
\item {\textbf{Baseline}} executes only on the CPU.
\item {\textbf{\xname{} SD}} offloads crash consistency operations to a single \xname{} device.
\item {\textbf{\xname{} MD SW-sync}}  offloads crash consistency operations to two \xname{} devices and synchronizes using a CPU-polling, software mechanism.
\item {\textbf{\xname{} MD}} offloads crash consistency operations to two  \xname{} devices with delayed synchronization. 
\end{itemize}

\subsection{Speedup Evaluation} 
In this section, we evaluate applications (listed in  \cref{tab:workloads}) in the configurations mentioned in \cref{subsec:methodology}.
We first demonstrate the benefit of NDP using a microbenchmark and demonstrate the parallelism in real applications. 
Then, we present the speedup of these applications of both crash consistency code regions and the whole programs. 

\begin{figure}
  \centering
 \includegraphics[width=.8\linewidth]{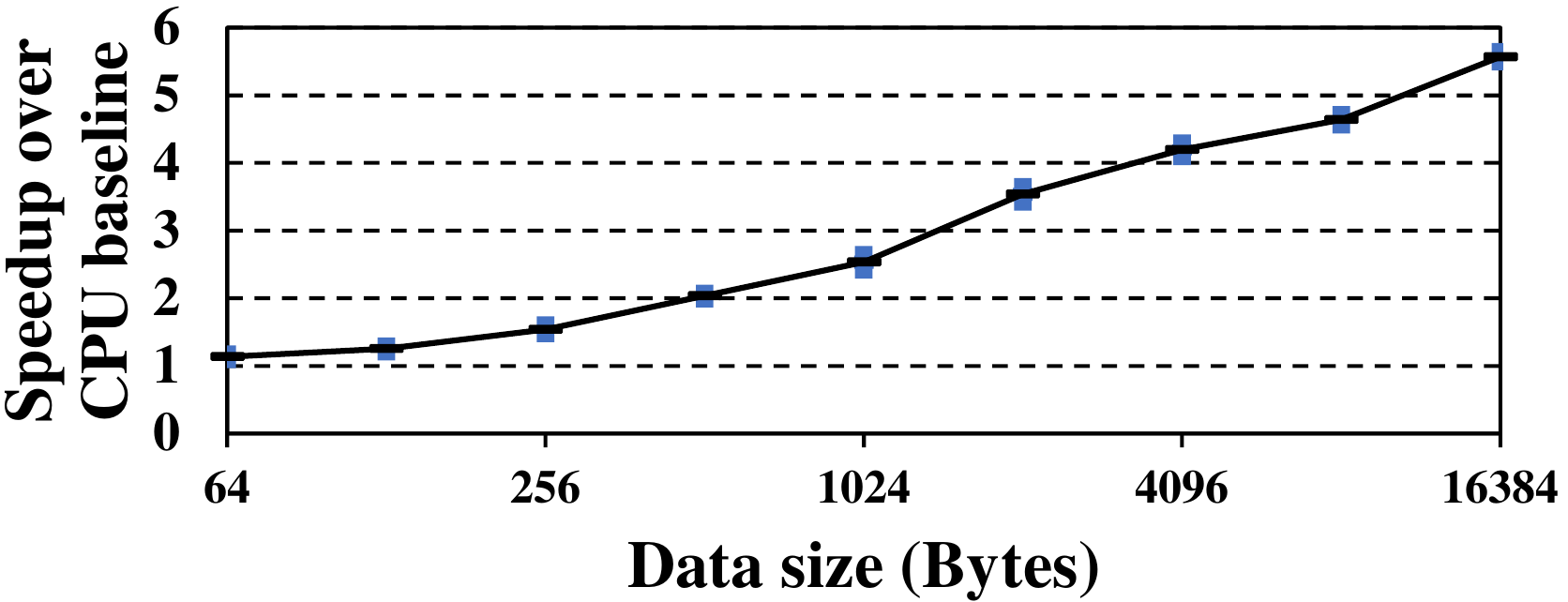}
  \caption{Data movement speedup of \xname{} over CPU.}\
  \label{fig:datacopy}
\end{figure}

\subsubsection{Micro-benchmark.} 

We evaluate \xname{} with a micro-benchmark that copies persistent data.
\cref{fig:datacopy} shows the speedup from \xname{}.
As data size increases, the speedup also increases: from $1.13\times$ when the size is 64 B to $5.57\times$  when copying 16 kB of data.
This micro-benchmark does not introduce operation-level parallelism. Thus, the speedup is a result of the proximity to memory when copying data with \xname{}. 
This speedup is comparable to prior FPGA-based NDP prototypes \cite{lee2021hardware,PiDRAM2021,upmem2021}.

\subsubsection{CPU-\xname{} parallelism.} \label{subsubsec:parallelism}
\begin{figure}
\centering
    \begin{minipage}[t]{0.45\linewidth}
    \centering
    \includegraphics[height=1.2in]{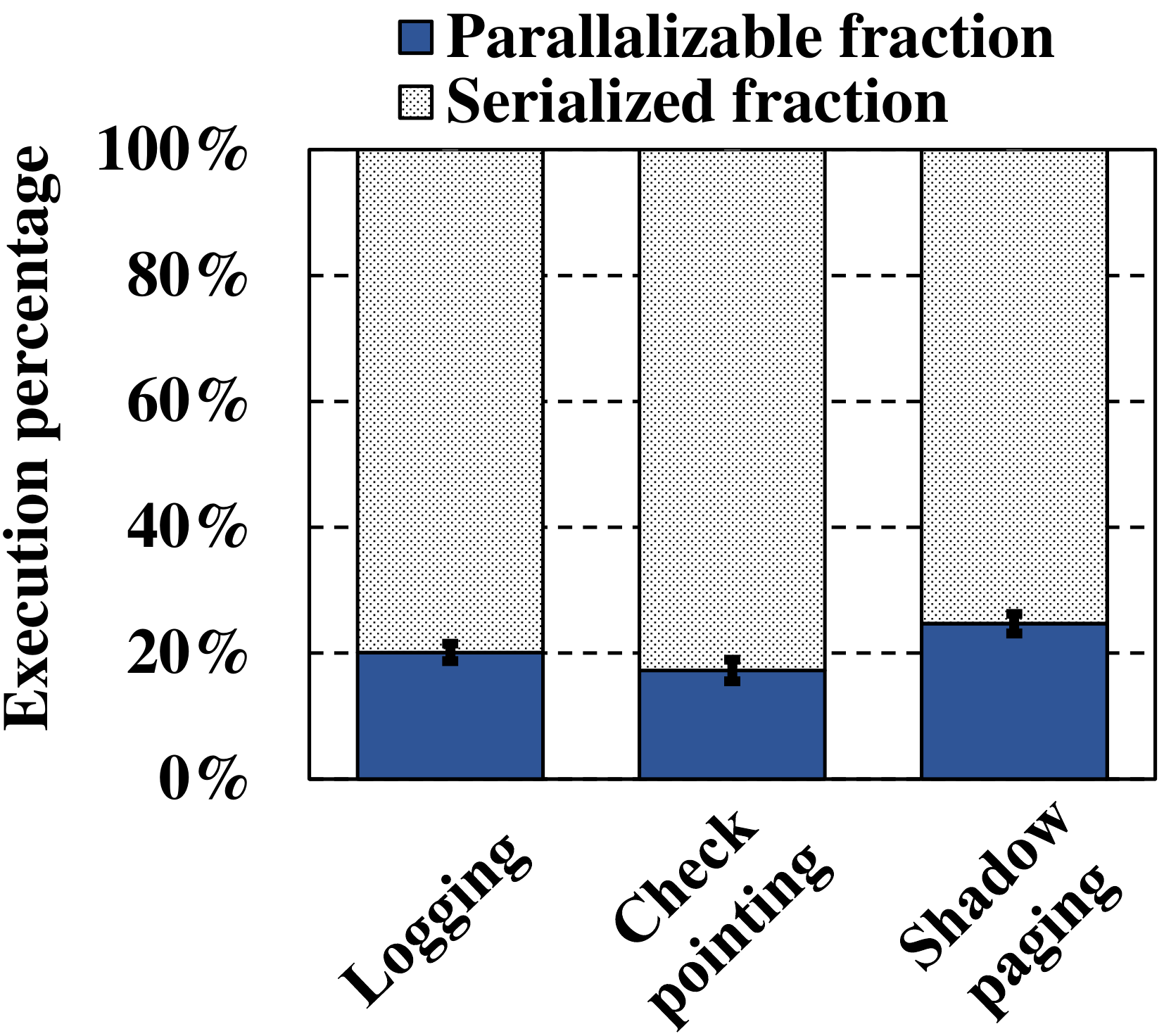}
    \caption{Percentage of parallel execution between CPU and \xname{}.}
    \label{fig:parallel}
    \end{minipage}
\hspace{0.05in}
    \begin{minipage}[t]{0.45\linewidth}
    \centering
    \includegraphics[height=1.2in]{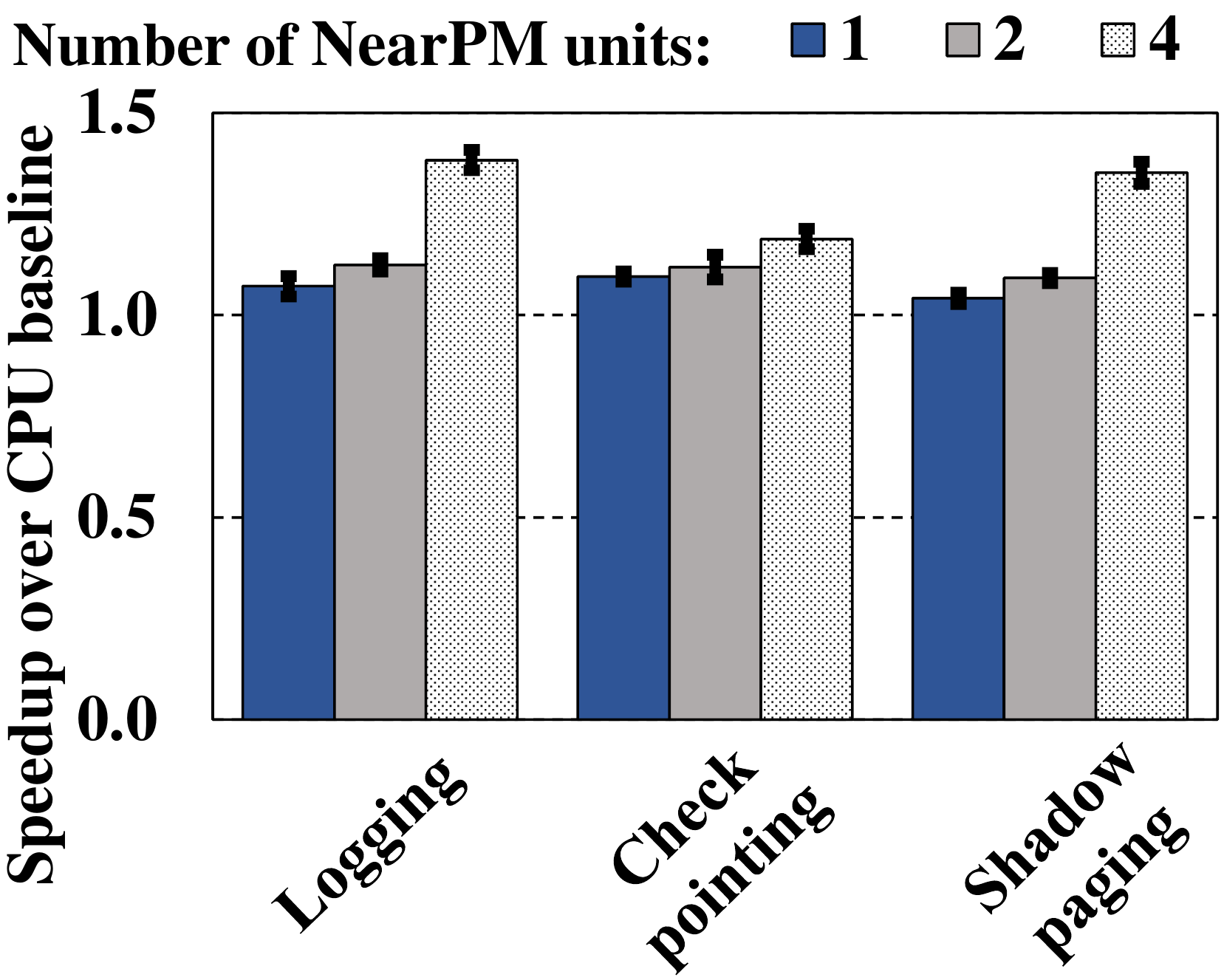}
    \caption{End-to-end performance with variable \# \xname{} Units.}
    \label{fig:numndp}
    \end{minipage}
\end{figure}

We evaluate the benefit of parallelism between CPU and  NDP devices, \ie CPU and \xname{} may execute at the same time for a certain fraction of the program.
\cref{fig:parallel} presents the average percentage of execution that is parallelizable  between the CPU and \xname{} of workloads  in \cref{tab:workloads}.
On average, logging, checkpointing, and shadow paging have 20.01\%, 17.25\%, and 24.68\% of the execution parallelizable, respectively. 

\subsubsection{Speedup in crash consistency operations.}
\cref{fig:crashp_kernel} shows the speedup within code regions that maintain crash consistency. 
On average \pname{} achieves $6.9\times$, $4.3\times$, and $9.8\times$ speedup for logging, checkpointing, and shadow paging, respectively.
We  notice that TATP has a low speedup of  $1.23\times$ in undo-logging. 
The main reason is that TATP has only one \xname{} operation that performs logging and commits immediately afterward. Thus, it does not benefit from parallelism in \xname{} execution.

{\sloppy
\subsubsection{Whole-application speedup.} \label{subsubsec:overall_perf}
We then present the whole-application performance in \cref{fig:crashp}.
\xname{} SD achieves $1.29\times$, $1.15\times$, and $1.28\times$ average speedup for logging, checkpointing, and shadow paging, respectively.
This result shows the performance \pname{} achieved by effective handling of ordering between the CPU and NDP.
\xname{} MD SW-sync achieves $1.21\times$, $1.14\times$, and $1.23\times$ average speedup for logging, checkpointing, and shadow paging, respectively. 
Due to the synchronization overhead, its speedup is lower compared to \xname{}~SD.
By reducing the synchronization overhead, \xname{}~MD achieves $1.35\times$, $1.22\times$, and $1.33\times$ speedup on average in the three crash consistency mechanisms, respectively. 
}

\subsection{Scalability Evaluation}

Next, we evaluate the scalability of \xname{}.
First, we demonstrate the multithreading performance of two realistic workloads, Memcached~\cite{pmem-memcached}, and Redis~\cite{pmem_redis}.
Then, we present the impact of the number of \xname{} units on performance, by sweeping the number of units.

\subsubsection{Multi-threaded Performance.}

\begin{figure}
  \centering
 \centering
    \includegraphics[height=1.3in]{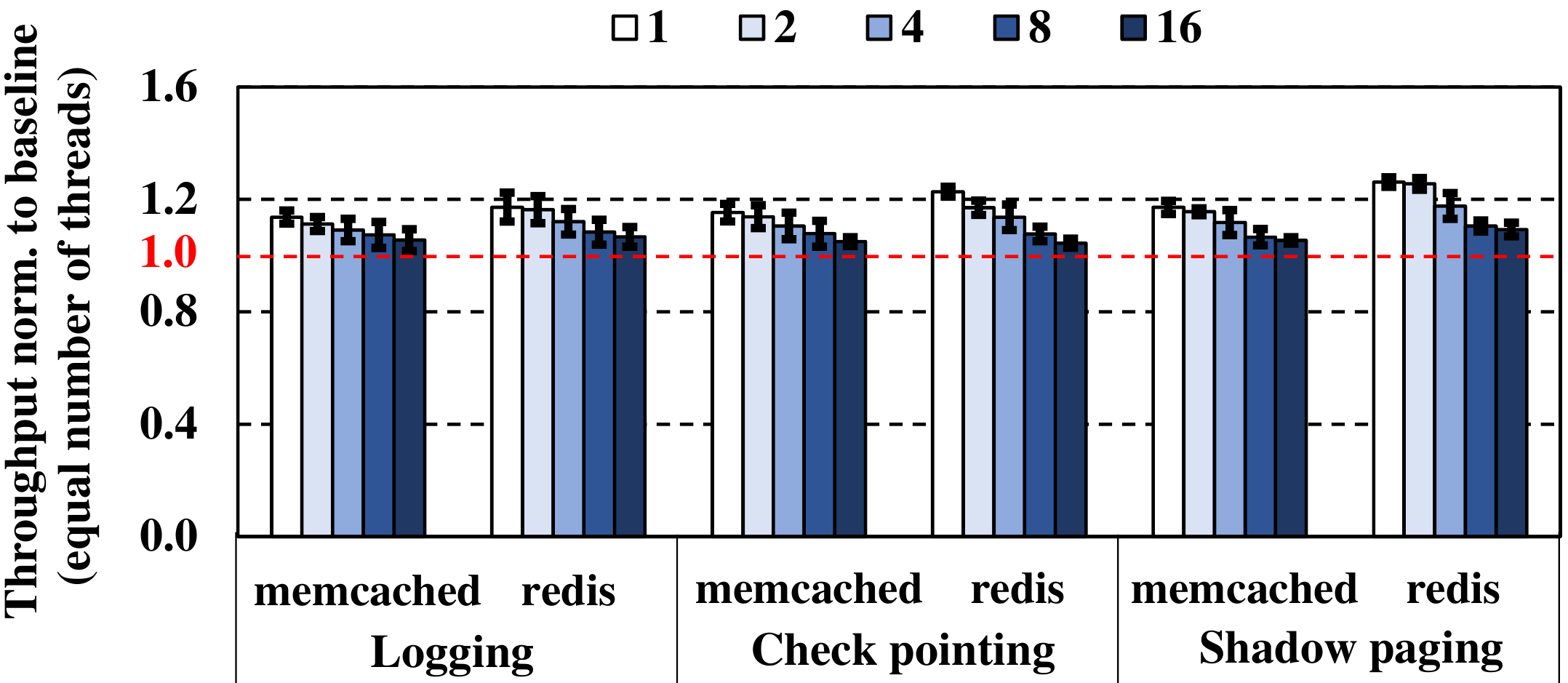}
    \caption{Throughput with multiple threads.}
    \label{fig:multi}
\end{figure}

This experiment evaluates the performance when the application on the host CPU is multithreaded.
We scale Redis and Memcached with 1 to 16 threads, for both the number of clients and the backend handlers.
Memcached access different memory pools for each thread, while Redis shares the same pool amongst multiple threads. 
\cref{fig:multi} presents the speedup over the CPU-based baseline with the same number of threads.
As the number of threads increases, the speedup from \xname{} reduces but still outperforms the baseline.
This trend still holds even when we consider standard error for multiple runs as shown in black in \cref{fig:multi}.
The main reason for the slowdown is that the number of execution units in \xname{} is limited to four due to the limitation of our FPGA platform. We expect commercialized systems to integrate more units for intensive workloads.

\subsubsection{Sensitivity study on the numbers of \xname{} units.}
This experiment compares the performance with 1, 2, and 4 \xname{} units.
\cref{fig:numndp} shows that the average speedup over the CPU-based baseline increases with more \xname{} units, as the offloaded program contains parallelizable operations, such as copying multiple cache lines in a page can happen in parallel. 

\section{Discussion}
\paragraph*{Scalability.}
Though in \cref{sec:eval}, we evaluated our prototype of two \xname{} devices, due to limitations in our FPGA platform, \pname{} is scalable as synchronizations among devices are off the critical path. 
Scalability is critical to performance with CXL-supported systems. 

\paragraph*{Expected performance in commercial NDP systems.}
Our \xname{} prototype shows comparable performance as prior work that also prototype NDP systems \cite{lee2021hardware,upmem2021,PiDRAM2021}---we achieve $7-9\times$ speed when evaluating the offloaded crash consistency operations (\ie accelerable computation kernels).
In commercialized implementations, we expect better performance as the \xname{} device can allow for more processing units that operate at a higher clock frequency. 

\paragraph*{Opportunity with CXL.}
As CXL is around the corner, future-generation PM systems are expected to be a CXL-based instead of occupying DIMM slots.
Though evaluated using a PCIe FPGA, \xname{} is independent of the interconnect technology and can largely benefit from the hardware-based coherence support from CXL. 
In the current design, the software handles coherence by explicitly writing updated data back to \xname{} devices. With CXL, we expect a much lower communication and synchronization overhead between NearPM devices and the host CPU.

\paragraph*{Security considerations.}
\xname{} target performance optimization using NDP while ensuring correctness. 
Thus, security in multi-tenant scenarios is not the focus. 
Nonetheless, to support multi-tenancy, the existing address-translation mechanism can be extended to support boundary checking by storing the pool size alongside the address translation offset.
We expect future work to build upon our prototype. 

\paragraph*{NUMA support.}
NUMA systems are common in datacenters. Although our evaluation platform is single-socket, the design of \xname{} fundamentally supports NUMA systems. 
The major challenge in NUMA system is that accesses to memory devices that belong to a different socket can experience a longer latency, increasing the variability of memory accesses. 
\xname{} guarantees a correct ordering of PM accesses and NDP-offloaded operations. Therefore, \xname{} is NUMA-safe.

\section{Related Work}

\paragraph*{Near-data processing.} NDP aims to reduce memory movement in the conventional CPU-centric systems \cite{ahn2015_ISCA,ahn2015_ISCA2,fernandez2020_ICCD,gao2015_PACT,gao2016_HPCA,hsieh2016_sigarch,hsieh2016_ICCD,kim2016_sigarch,mutlu2020modern,singh2020_FPL,singh2019_DAC,zhan16_micro,kim2018_BMC,gaofracdram,gaocompdram}.
For example, RowClone~\cite{seshadri2013isca} accelerates bulk data movement in DRAM and TETRIS~\cite{gao2017asplos} accelerates neural networks. 
There have also been works that bring processing to SSDs. 
For example, CompoundFS \cite{ren20_compoundfs}  accelerates file system IO operations in SSD and Almanac \cite{wang19_almanac} retains SSD history logs using an in-SSD logic. 
However, they target conventional storage systems instead of PM systems that directly manage persistent data.

\paragraph*{Hardware support for memory persistency.}
The memory persistency model ensures the order in which writes become persistent. 
Pelly et al. first propose memory persistency \cite{pelley14_isca} and followup works continue to optimize the performance of persistency models. 
For example, DPO \cite{kolli16_micro} and  HOPS \cite{nalli17_asplos}, PMEM-Spec \cite{jeong21_pmemspec}, and Themis\cite{Shahri2020Fenceless} provide  efficient persistency models by reducing the cost of blocking due to data persistence. 
However, those works target CPU-centric systems. 
In comparison, our work, \xname{}, extends persistence to NDP.

\paragraph*{Crash consistency mechanisms.}
There are a number of previous works that provide solutions for crash consistency.
Intel's PMDK library provides transactions using a combination of undo and redo logs~\cite{libpmemobj}.
There are also databases and key-value stores based on PMDK that maintain crash consistency, such as Redis \cite{pmem_redis}, MongoDB \cite{pmem_mongodb}, RocksDB \cite{pmem_rocksdb}, and Memcached \cite{pmem-memcached}.
Atlas \cite{chakrabarti14_oopsla} and SFR~\cite{gogte18_pldi} convert code regions marked by synchronization primitives to undo-log-based transactions.
Checkpointing creates a copy of the updated persistent memory to enable recovery \cite{dong2009SC, joshi15_micro}.
DudeTM \cite{liu17_asplos} and SoftWrAP \cite{giles15_msst} use shadow memory to maintain redo logs before applying them to PM. 
These mechanisms tend to maintain additional copies of data for recovery.
Therefore, \xname{}, can be applied to mitigate their crash consistency overhead.
Recently works also use software testing techniques to detect bugs in crash-consistent programs \cite{liu19_asplos, pmdebugger, pmfuzz, hippo, jaaru}. These techniques can also be applied to a \xname{} system to detect misuse of offloaded crash consistency primitives. 
\section{Conclusions}
In this work, we propose \xname{}, an accelerator for crash consistency mechanisms in PM-based storage-class applications. 
To realize the full potential of acceleration, we move the ordering handling between CPU and \xname{} near memory, using Partitioned Persist Ordering (\pname{}).
We prototype \xname{} on an FPGA platform and evaluate nine PM workloads, where each workload has three versions that use logging, checkpointing, and shadow paging for crash consistency. 
Overall, \xname{} achieves $4.3-9.8\times$ speedup in the NDP-offloaded operations and $1.22-1.35\times$ speedup in end-to-end execution of the whole applications.

\section*{Acknowledgement}

We thank the anonymous reviewers and the shepherd, Prof.~Marc Shapiro, for their valuable feedback.
This work is supported by the SRC/DARPA Center for Research on Intelligent Storage and Processing-in-memory (CRISP) and the National Science Foundation (NSF).

\bibliographystyle{plain}
\bibliography{bibtex/ref}
%

\appendix
\section{Artifact Appendix} 

\subsection{Abstract}
This artifact document consists of the description of how to reproduce the major results of the \xname{} accelerator evaluation results in \cref{sec:eval}.
The evaluation is performed on a Xilinx Virtex UltraScale+ VCU118 Evaluation Platform, where the on-board DRAM is emulated as PM, accessible by both the host CPU and NDP units in \xname{}.

\subsection{Description \& Requirements}

\subsubsection{How to access}

The source code of the software and the FPGA implementation are hosted on two GitHub repositories:
\begin{itemize}
    \item \url{https://github.com/Systems-ShiftLab/NearPMSW}
    \item \url{https://github.com/Systems-ShiftLab/NearPMHW}
\end{itemize}

\subsubsection{Hardware dependencies}
\begin{itemize}
    \item CPU: AMD Ryzen 7 3700X CPU
    \item Memory: 64 GB DDR4
    \item FPGA: Xilinx Virtex UltraScale+ VCU118 evaluation platform
\end{itemize}

The hardware build process uses the Vivado toolchain (version 2018.2 in our experiments). 
A licensed Vivado version is required to create the bitstreams for the experiments in this document.
The toolchain compiles and deploys the bitstreams.

\subsubsection{Software dependencies}
The software dependencies include both kernel modifications and PM workloads. 
First, the kernel of the system must be modified and recompiled to enable caching for the emulated PM. 
The instructions for the required kernel change are provided in \path{NearPMHW/kernelchange.pdf}.
In our setup, the evaluation was done using Linux kernel version 5.4.0. 
The workloads in the software repository \path{NearPMSW} have their own dependencies, which are included in our repository.

\subsection{Set-up}
\label{sec:setup}
Clone the NearPMHW repository and create the Vivado project as  \path{NearPMHW/NearPM}:
\begin{lstlisting}[basicstyle=\scriptsize]
$ git clone https://github.com/Systems-ShiftLab/NearPMHW
$ cd NearPMHW
$ vivado -mode batch -source build.tcl
\end{lstlisting}

\noindent Next, launch Vivado in GUI mode.
\begin{lstlisting}[basicstyle=\scriptsize]
$ vivado 
\end{lstlisting}

\noindent Click on \emph{Open Project} and open the newly created NearPM project.
After the project loading is completed you will find out five designs on the \emph{Sources} tab. 
To build any of the designs: First, right-click on the design and select ``Set as Top''. Next, click ``Generate Bitstream'' in ``Flow Navigator'' tab and follow the prompt.

This process will take approximately one hour. After the bitstream is generated, program the FPGA using the ``Hardware Manager'' (bottom left of the Vivado window).

First, make sure that caching of the FPGA memory region is functional. 
Generate the bitstream for design5 in the project. Use the previously explained steps to generate the bitstream. 
After programming the device, restart the host and then run the following commands for testing.
\begin{lstlisting}[basicstyle=\scriptsize]
$ source NearPMHW/setup.sh
$ git clone https://github.com/Systems-ShiftLab/NearPMSW
$ cd NearPMSW/pcache
$ make 
$ sudo ./random-chase
\end{lstlisting}

If caching works properly, the access latency will be around  1--2~ns for smaller block sizes due to caching and keeps increasing as the program proceeds due to of cache misses when accessing larger blocks. 
You may terminate the test program after analyzing several data sizes.
\subsection{Evaluation workflow}

\subsubsection{Major Claims}
The artifact comes with scripts that reproduce the key results in the following figures:

\begin{itemize}
    \item \textbf{\cref{fig:crashp_kernel}:} Speedup in code regions for crash consistency in (a) logging, (b) checkpointing, and (c) shadow paging.
    \item \textbf{\cref{fig:crashp}:} End-to-end speedup in (a) logging, (b) checkpointing, and (c) shadow paging.
\end{itemize}

\subsubsection{Experiments}
After setting up the environment, reproduce the following experiments. 

\textit{\textbf{Experiment (E1):} [\cref{fig:crashp_kernel}] [60 human-minutes + 2 compute-hour]: The experiment will reproduce the results in \cref{fig:crashp_kernel}.}\\

For this result to be regenerated the generated bitstream of design5 of the Vivado project must be loaded onto the FPGA.
Restart the host and run the following commands. 
\begin{lstlisting}[basicstyle=\scriptsize]
$ cd NearPMSW
$ ./genfig15.sh
\end{lstlisting}

The output will present the results in a readable format.

\textit{\textbf{Experiment (E2):} [\cref{fig:crashp}] [60 human-minutes + 2 compute-hour]: The experiment will reproduce the results in \cref{fig:crashp}.}\\

For this result to be regenerated the generated bitstream of design5 of the Vivado project must be first loaded onto the FPGA.
Restart the host and run the following commands. 
\begin{lstlisting}[basicstyle=\scriptsize]
  $ source ./NearPMHW/setup.sh 
\end{lstlisting}
Next, run the following commands to get the results for the \xname{} MD-sync case.
\begin{lstlisting}[basicstyle=\scriptsize]
$ cd NearPMSW
$ ./genMDsync.sh
\end{lstlisting}

\noindent Next, take the following steps on Vivado: 
\begin{enumerate}
    \item Open Vivado GUI. 
    \item Click on \emph{Open Block Design} and select \emph{design\_5}.
    \item Double click on the block \emph{multi\_thread\_command\_0}.
    \item Change \emph{Dimm0 End Addr} to \texttt{0xBFFFFFFF}.
    \item Follow the steps in \cref{sec:setup} to generate bitstream and program the FPGA.
\end{enumerate}

\noindent After restarting the host, run the following commands.
\begin{lstlisting}[basicstyle=\scriptsize]
$ source NearPMHW/setup.sh
$ cd NearPMSW
$ ./genfig16.sh
\end{lstlisting}

\end{document}